\documentclass[a4paper, 12pt]{article}
	
\usepackage{natbib}
	\setcitestyle{authoryear,round,aysep={}}
\usepackage{subfigure}
\usepackage{float}
\usepackage{amssymb}
\usepackage{authblk}
\usepackage{graphicx}
\usepackage{geometry}
\usepackage{enumitem}
\usepackage{amsmath}
\usepackage{url}
\usepackage[utf8]{inputenc}
\DeclareUnicodeCharacter{2032}{}
\usepackage{microtype}
\usepackage[hidelinks]{hyperref}
\title{Modelling Socio-Psychological Drivers of Land Management Intensity}

\author[1]{Ronja Hotz}
\author[1]{Calum Brown}
\author[1]{Yongchao Zeng}
\author[1]{Thomas Schmitt}
\author[1,2,3]{Mark Rounsevell}

\affil[1]{Institute of Meteorology and Climate Research, Atmospheric Environmental Research (IMK-IFU), Karlsruhe Institute of Technology, 82467 Garmisch-Partenkirchen, Germany}
\affil[2]{Institute of Geography and Geo-Ecology, Karlsruhe Institute of Technology, 76131 Karlsruhe, Germany}
\affil[3]{School of Geosciences, University of Edinburgh, Drummond Street, Edinburgh EH8 9XP, UK}

\date{}

\begin{document}

\maketitle

\begin{abstract}
Land management intensity shapes ecosystem service provision, socio-ecological resilience and is central to sustainable transformation. Yet most land use models emphasise economic and biophysical drivers, while socio-psychological factors influencing land managers’ decisions remain underrepresented despite increasing evidence that they shape land management choices. To address this gap, we develop a generic behavioural extension for agent-based land use models, guided by the Theory of Planned Behaviour as an overarching conceptual framework. The extension integrates environmental attitudes, descriptive social norms and behavioural inertia into land managers’ decisions on land management intensity. To demonstrate applicability, the extension is coupled to an existing land use modelling framework and explored in stylised settings to isolate behavioural mechanisms.
Results show that socio-psychological drivers can significantly alter land management intensity shares, landscape configuration, and ecosystem service provision. Nonlinear feedbacks between these drivers, spatial resource heterogeneity, and ecosystem service demand lead to emergent dynamics that are sometimes counter-intuitive and can diverge from the agent-level decision rules. Increasing the influence of social norms generates spatial clustering and higher landscape connectivity, while feedbacks between behavioural factors can lead to path dependence, lock-in effects, and the emergence of multiple stable regimes with sharp transitions.
The proposed framework demonstrates how even low levels of behavioural diversity and social interactions can reshape system-level land use outcomes and provides a reusable modelling component for incorporating socio-psychological processes into land use simulations. The approach can be integrated into other agent-based land use models and parameterised empirically in future work.
\end{abstract}


\section{Introduction}

With a growing human population, the global demand for food, timber, and other ecosystem services (ES) continues to increase \citep{diaz_pervasive_2019}. At the same time, land use change is one of the main drivers of environmental degradation \citep{foley_global_2005, ipbes_summary_2018}. Specifically, high land management intensity is a key factor contributing to the transgression of multiple planetary boundaries, including freshwater use, nitrogen and phosphorus cycles, climate change, and the integrity of the biosphere \citep{campbell_agriculture_2017, rockstrom_safe_2023,ipbes_summary_2019, mcelwee_ipbes_2025}. This makes transitions to more environmentally sustainable land management practices a priority.
Understanding how such transitions unfold requires close attention to how land managers make decisions about land management intensity.
These decisions are often context-specific and influenced by various interacting factors \citep{swart_meta-analyses_2023,brown_simplistic_2021,dessart_behavioural_2019}.
While economic, structural and natural conditions are well documented in the scientific literature, socio-psychological factors remain less studied, despite growing evidence of their importance for land use decisions \citep{swart_meta-analyses_2023, dessart_behavioural_2019,burton_good_2020}.
Socio-psychological factors refer to the cognitive, social and emotional influences on behaviour including individuals' values, attitudes, norms, and beliefs, and significantly shape land management intensity choices such as the adoption of environmentally sustainable management practices \citep{ maybery_categorising_2005, franzini_key_2024, diana_feliciano_decision_2025, westin_forest_2023, brown_simplistic_2021}.

Land managers' decisions are embedded within their social context \citep{klebl_farmers_2024, sotirov_forest_2019}, often shaped by peers, family, and trusted advisors \citep{dessart_behavioural_2019, brown_simplistic_2021}. The behaviour of neighbouring land managers is particularly influential in determining the adoption of sustainable land use practises \citep{dessart_behavioural_2019, schmidtner_spatial_2012, lapple_spatial_2015}. For instance, farmers are less likely to adopt agri-environmental measures if neighbouring farmers have limited experience with such practises. Conversely, awareness that sustainable practises are widely adopted within their community positively influences their adoption \citep{dessart_behavioural_2019}. Similarly, forestry practitioners across Europe place considerable importance on social norms and behaviours exhibited by neighbouring land users \citep{diana_feliciano_decision_2025}, underscoring how social networks and peer behaviour influence their land management practises.

Alongside social influences, land managers' individual values and beliefs critically affect their land use decisions \citep{brown_simplistic_2021,diana_feliciano_decision_2025}. Land managers often hold multiple, sometimes conflicting, values simultaneously—such as maximising production and income, preserving the environment, or maintaining cultural traditions \citep{gosling_connectedness_2010, westin_forest_2023, thompson_spirit_2017, klebl_farmers_2024, thompson_farmers_2024, huttunen_agri-environmental_2016}. Generally, strong environmental values can direct management practises towards conservation and sustainability, while economically-oriented values can steer decisions towards increased productivity \citep{brown_simplistic_2021}. For instance, farmers’ pro-environmental attitudes can be positively correlated with participation in biodiversity schemes, while productivist attitudes can be negatively correlated \citep{breustedt_factors_2013, espinosa-goded_identifying_2013,grammatikopoulou_locally_2013,kvakkestad_norwegian_2015,micha_uptake_2015}.
Likewise in forestry, environmental values tend to guide management practises towards nature conservation, such as deadwood retention and selective cutting \citep{franzini_key_2024, husa_non-industrial_2021,joa_conservation_2020, vedel_forest_2015}.

Despite these well-documented behavioural dynamics, most large-scale land use models — and many policy instruments such as the EU’s Common Agricultural Policy — continue to emphasise economic optimisation and biophysical constraints \citep{dessart_behavioural_2019}. To anticipate, manage, and support sustainable land use transitions, modelling land use change is essential. Models allow researchers and policymakers to explore potential future scenarios, test policy interventions, and understand how different decision-making processes aggregate to produce system-level outcomes \citep{murray-rust_open_2014, murray-rust_combining_2014}.
Traditional models — including macroeconomic simulations, integrated assessment models, and Earth system models — offer valuable insights into global land use dynamics but often overlook the complex decision-making processes of individual land managers.
Agent-based land use models (ABLUMs) offer a more flexible framework, allowing for heterogeneity and interaction among decision-makers \citep{murray-rust_combining_2014}. However, even many ABLUMs remain rooted in economic rationality and still lack robust representations of socio-psychological drivers such as values, norms, and social interactions \citep{huber_representation_2018, groeneveld_theoretical_2017}.

To address this gap, we develop a reusable behavioural model extension for ABLUMS that explicitly incorporates socio-psychological factors in land management intensity decisions. Drawing on the Theory of Planned Behaviour \citep{ajzen_theory_1991,ajzen_intentions_1985} as an overarching framework, the extension integrates environmental attitudes, descriptive social norms, and behavioural inertia as a proxy for limited perceived behavioural control. The decision-making model is conceptually generic and designed to be transferable across ABLUMs with minor refinements. We integrate the extension into an existing agent-based land use modelling framework and explore it in stylised settings to isolate behavioural dynamics from confounding real-world complexity. With behavioural factors we refer to all factors shaping land managers' decision-making and behaviour.
Specifically, we investigate how socio-psychological drivers interact with biophysical resource distribution and ES demand to produce emergent system outcomes. We therefore aim to answer the following research questions:
\begin{enumerate}
\item How does the integration of socio-psychological factors into land management intensity decisions alter emergent land management intensity patterns and total supply of ES?
\item What system-level dynamics (asymmetry, clustering, critical transitions, path dependence) arise from interacting socio-psychological, economic and biophysical factors shaping land management intensity decisions?
\end{enumerate}

\section{Methods}
\subsection{Model Description}
\begin{figure}[!t]
\centering
\includegraphics[width=0.8\linewidth]{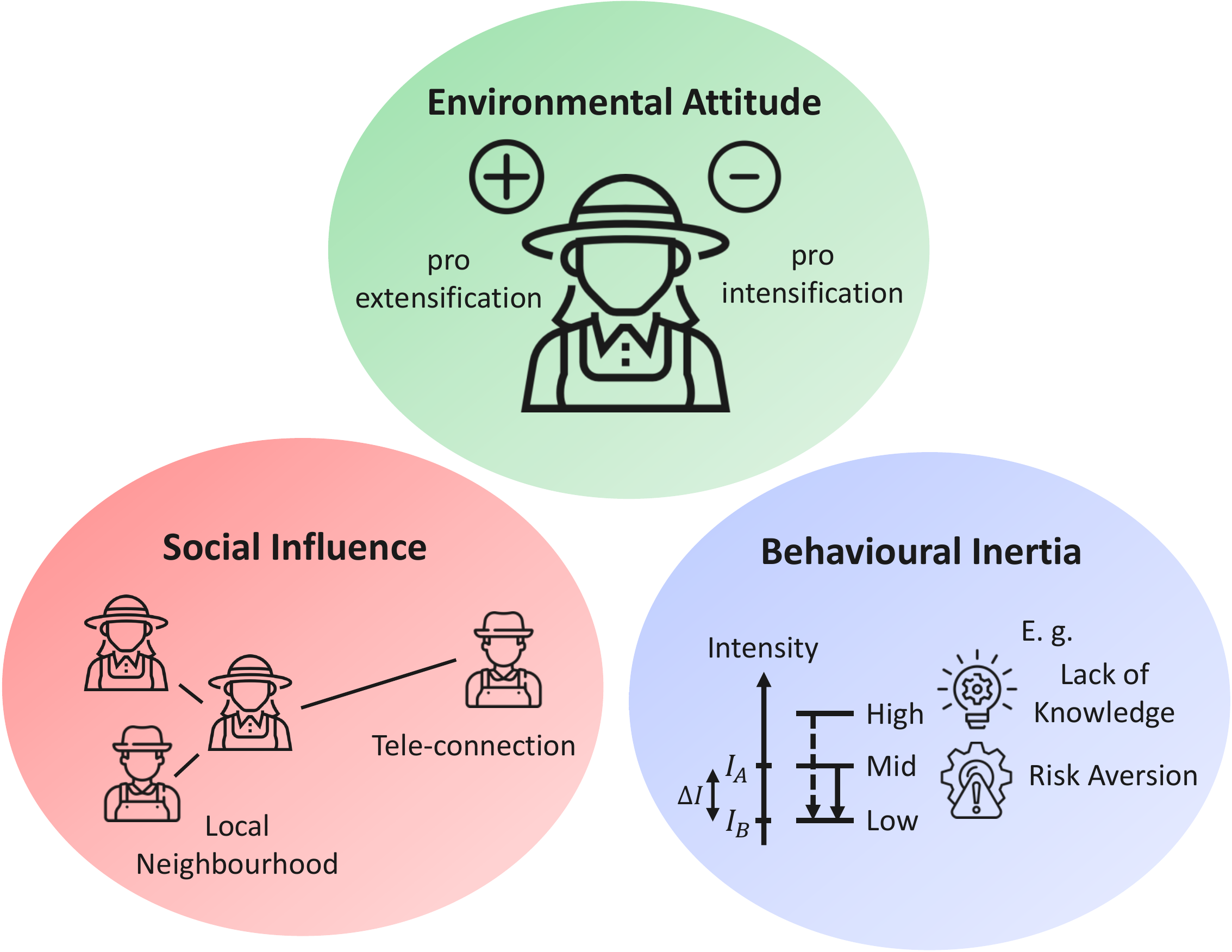}
\caption{Visualisation of the different behavioural factors included in the model extension.}
\label{fig:behavioural-factors}
\end{figure}
We present a generic model extension for land managers' decisions on changing land management intensity, using the Theory of Planned Behaviour (TPB) \citep{ajzen_theory_1991} as a guiding framework to complement existing behavioural mechanisms such as economic maximization in ABLUMs. The TPB is a well-established and empirically grounded framework in cognitive psychology that has been widely applied in various contexts including land use \citep{kaufmann_simulating_2009, borges_using_2016,daxini_understanding_2019,vaz_identifying_2020}.
According to the TPB, human behaviour is driven by three core components: attitudes, subjective norms and perceived behavioural control \citep{ajzen_theory_1991}. 
Mapped to the land use context, we conceptualise attitudes as environmental attitudes, subjective norms as descriptive social norms and limited perceived behavioural control as behavioural inertia, visualised by Figure \ref{fig:behavioural-factors}. This modelling choice is further explained in Subsection \ref{subsubsec:attitudes}-\ref{subsubsec:inertia}. 
In operational terms, the behavioural extension provides a decision layer that determines whether land managers switch between different land management intensities based on weighted environmental attitudes, descriptive social norms, and behavioural inertia. This is implemented through a giving-in threshold that must be exceeded for a behavioural change to occur, which can interact with biophysical constraints and economic competition when the extension is embedded within an ABLUM. While designed as a threshold-based mechanism, the logic can be reformulated into probabilistic adoption functions for models that operate with stochastic decision structures. Conceptually, the extension functions as a generic behavioural module that translates socio-psychological drivers into land management decisions while remaining independent of the specific land-use allocation processes of the host model. Environmental attitudes bias managers toward either extensification or intensification depending on their underlying values. Descriptive social norms capture peer influence within a social network that combines spatial neighbourhoods with optional long-distance tele-connections. Behavioural inertia reflects limited perceived control and increases with the magnitude of the intended management change, thereby raising the threshold for switching intensity levels and stabilising current behaviour.
A flow diagram of the behavioural model can be found in the Appendix (Figure \ref{fig:flow_diagramm}).

\subsubsection{Environmental Attitude}
\label{subsubsec:attitudes}

According to a recent meta-analysis, attitude is consistently identified as one of the strongest predictors of sustainable agriculture adoption among farmers in Europe \citep{swart_meta-analyses_2023}. Land managers’ attitudes towards changing land management intensity are shaped by economic, environmental, and social motivations \citep{klebl_farmers_2024, westin_forest_2023}. We specifically represent attitudes as environmental attitudes here in order to complement economic theories that are already central in ABLUMs \citep{groeneveld_theoretical_2017}. 

In our model, the environmental attitude $A_{\alpha}$ represents the land manager’s preference for land use  extensification or intensification guided by environmental or productivist values, respectively. We generally denote by $\alpha$ the index referring to land manager-specific behavioural characteristics. 

Following empirical studies, land managers' pro-environmental attitudes are often positively associated with the adoption of less intensive and more conservation-oriented land management practices, whereas economic-oriented attitudes tend to steer management towards increased production \citep{breustedt_factors_2013, espinosa-goded_identifying_2013, grammatikopoulou_locally_2013, kvakkestad_norwegian_2015, micha_uptake_2015}. We build on this relationship by assuming that land managers who hold pro-environmental attitudes, represented by $A_{\alpha}>0$, are more inclined toward extensification. Specifically, we implement this by lowering the behavioural threshold for adopting more extensive land use practices and raising it for intensification, see Subsection \ref{subsubsec:behavioural_function}. The inverse model mechanism is implemented for land managers holding productivist attitudes, represented by $A_{\alpha}<0$.

 Formally, the effect of the land manager's environmental attitude on a considered transition from current land mangement intensity $I_A$ to prospective land management intensity $I_B$ is given by:
\begin{equation}
A_{AB}^{\alpha} = -\text{sign}(I_B - I_A) \cdot A_{\alpha},
\end{equation}
where $\text{sign}(\cdot)$ denotes the sign function. This formulation adjusts the influence of $A_{\alpha}$ depending on the direction of change:
\begin{equation}
A_{AB}^{\alpha} =
\begin{cases}
+ A_{\alpha}, & \text{if } I_B < I_A \text{ (extensification)}, \\
- A_{\alpha}, & \text{if } I_B > I_A \text{ (intensification)}.
\end{cases}
\end{equation}
That is, a pro-environmental attitude exerts a positive influence on decisions favouring extensification, and a negative influence on decisions favouring intensification, whereas a productivist attitude has the opposite effect. These attitudinal tendencies do not automatically translate into the corresponding behavioural responses: holding pro-environmental attitudes does not guarantee the adoption of environmentally sustainable practices in our model, as behaviour may still be constrained or counteracted by social influence, economic pressure, or other model components.

\subsubsection{Descriptive Social Norm}
\label{subsubsec:descriptive_norms}

We conceptualize social norms as descriptive social norms, referring to individuals’ perceptions of how most others behave in a given context \citep{cialdini_focus_1990}. Descriptive social norms play a significant role in shaping land managers’ decisions. For example, \citet{dessart_behavioural_2019} found that farmers’ adoption of sustainable agricultural practices in Europe is significantly influenced by their perceptions of what other farmers in their region are doing. Complementing this, \citet{conley_learning_2010} provide evidence that farmers adjust their input use based on the observed outcomes of peers within their social network. Similarly,  \citet{brown_empirical_2018} show  evidence of spatial diffusion in the uptake of agricultural and forestry practices in the UK, suggesting that land use decisions are shaped by local social interactions. This importance of local peer influence is further supported by spatial econometric analyses and qualitative studies \citep{schmidtner_spatial_2012, lapple_spatial_2015, eastwood_cup_2022}.

To operationalise these insights, we adopt Bicchieri’s theory \citep{bicchieri_norms_2016}, in which behaviour is shaped by empirical expectations corresponding to what individuals believe others are doing. Each individual holds a ‘norm sensitivity’, reflecting the proportion of neighbours who must exhibit a behaviour before the individual considers adopting it. 
 In our model, each land manager observes the land use intensities of its neighbours within a social network and updates its expectations accordingly. 
 
 The social network is initially structured as a two-dimensional lattice, where each agent is connected to its spatial neighbours within a user-defined Moore radius.
This allows the model user to flexibly adjust the scope of social observation, reflecting differences in how locally embedded or spatially extended social influence may be in different land use contexts.

To incorporate broader social connectivity beyond strict spatial proximity, the network can be augmented using the Watts–Strogatz algorithm \citep{watts_collective_1998}. Starting from the spatial lattice, a user-defined amount of additional ties is build to randomly selected agents, creating occasional long-range links. This results in a network with small-world properties, characterised by high clustering and short average path lengths, which are often observed in real-world social networks \citep{watts_collective_1998,newman_structure_2003}. 

When an agent considers adopting a candidate land management intensity in our model, it forms an empirical expectation based on the prevalence of this land management intensity in its neighbourhood. If the candidate land use practice is more intensive than the current one ($I_B > I_A$), the agent evaluates the share of neighbours exhibiting the same or higher intensity. Conversely, for extensification ($I_B < I_A$), it considers the share of neighbours practising at or below the prospective level. This results in the following social influence score (referred to as net social pressure in Figure \ref{fig:flow_diagramm}):
\begin{align}
S_{AB}^{\alpha} =
\begin{cases}
p_{N^{\alpha}_A}(I \geq I_B) - CM_{int,\alpha}, & \text{if } I_B > I_A, \\
p_{N^{\alpha}_A}(I \leq I_B) - CM_{ext,\alpha}, & \text{if } I_B < I_A.
\end{cases}
\end{align}
Here, $p_{N^{\alpha}_A}(\cdot)$ denotes the proportion of neighbours in the agent’s social network $N^{\alpha}_A$ who meet the behavioural condition, and $CM_{int,\alpha}$ and $CM_{ext,\alpha}$ represent the agent's critical mass threshold for intensification and extensification, respectively. The concept of critical mass is commonly used in innovation diffusion \citep{rogers_diffusion_1983} and captures the agent’s norm sensitivity, i.e. the level of observed peer conformity required before the agent considers changing behaviour. Our model represents an adaptation of Granovetter’s threshold model of collective behaviour \citep{granovetter_threshold_1978}. It extends the original framework to a context in which agents choose between ordinal behavioural options (intensity levels) and replaces the hard threshold condition with a smooth threshold function, as described in the behavioural influence function (Subsection \ref{subsubsec:behavioural_function}). An agent perceives increased social pressure to adopt the candidate behaviour when its empirical expectation meets or exceeds its individual critical mass threshold. This structure enables us to model both an increase or a decrease in intensity with potentially different critical mass thresholds.

\subsubsection{Behavioural Inertia}
\label{subsubsec:inertia}

Within the TPB, behavioural change depends on the perceived behavioural control which refers to an individual’s perception of their ability to perform a given behaviour \citep{ajzen_theory_1991}. While actual behavioural control refers to the objective availability of resources or opportunities, \citet{ajzen_theory_1991} notes that perceived control is often more predictive of action, particularly when external constraints are ambiguous. Many agent-based land use models (ABLUMs), however, assume near-perfect knowledge and rationality \citep{groeneveld_theoretical_2017}, thereby overestimating behavioural responsiveness and overlooking cognitive or psychological constraints. In land use contexts, even when external conditions technically allow for change, land managers may still perceive the adoption of new practices as risky, complex, or personally unfeasible. Contributing factors include limited knowledge, uncertainty about future outcomes, low self-efficacy, or risk aversion—particularly where land use decisions involve delayed rewards or high stakes \citep{dessart_behavioural_2019}. If land managers perceive great difficulties to adopt new land use practices, they are more hesitant to adopt them especially when big changes in intensity are involved \citep{barreiro-hurle_does_2010}. 

To account for this gap between actual and perceived control, we introduce behavioural inertia into our model. Inertia represents a general reluctance to change, modelling the internal resistance agents may experience even when economic signals, personal values and social norms favour a shift in behaviour. Specifically, it enters the decision process as an additive resistance term that must be overcome by the behavioural drivers incorporated in the model, see Equation \ref{eq:x}. Behavioural inertia thus serves as a proxy for reduced perceived behavioural control, allowing us to better reflect realistic barriers to behavioural change in land management. 

Inertia is modelled as a function of the magnitude of the considered change in land management intensity: the greater the difference between the current intensity $I_A$ and the candidate intensity $I_B$, the stronger the behavioural resistance. This allows the model to reflect that large behavioural shifts are often perceived as more difficult or risky, while smaller adjustments are more likely to be considered feasible. The scaling of inertia with the size of the proposed change is further weighted by an agent-specific inertia coefficient $\lambda_{\alpha}$, capturing individual variation in behavioural conservatism. This formulation aligns with empirical evidence that land use change tends to occur incrementally rather than through abrupt transitions \citep{lambin_land_2010, van_vliet_manifestations_2015}. Larger shifts in management practices typically entail higher transaction costs and greater adaptation efforts, contributing to a preference for gradual behavioural change.
\subsubsection{Behavioural Influence Function}
\label{subsubsec:behavioural_function}

The decision to adopt a new land management intensity is governed by a behavioural threshold that reflects the combined influence of attitudinal, social, and inertia-related factors. We define this giving-in threshold $GIT_{AB}^{\alpha}$ using a logistic function that translates the aggregated behavioural drivers into a bounded decision score:
\begin{align}
GIT_{AB}^{\alpha} = \frac{L_{\alpha}}{1 + e^{k \cdot x_{AB}^{\alpha}}},
\label{eq:GIT}
\end{align}
where \( x_{AB}^{\alpha} \) is the composite behavioural influence score, \( L_{\alpha} \) denotes the upper bound of the threshold range, and \( k \) is a steepness parameter controlling the curvature of the logistic function.

This formulation ensures that the threshold remains within a well-defined range, supporting integration with other model components such as economic considerations or environmental constraints. The S-shaped profile of the logistic function captures key characteristics of real-world behavioural adoption dynamics—slow initial response, rapid transition near a critical point, and saturation at high influence levels  \citep{rogers_diffusion_1983}. It is widely used in the social sciences to model binary or probabilistic decisions \citep{harrell__regression_2015}, and in this context, it allows for flexible behavioural responses. Human decision-making often follows complex, non-proportional patterns that are captured more accurately by non-linear models such as the logistic function \citep{kim_human_2008}. 

The behavioural influence score \( x_{AB}^{\alpha} \) aggregates three key components: the influence of descriptive social norms \( S_{AB}^{\alpha} \), the attitudinal alignment with the proposed behavioural change \( A_{AB}^{\alpha} \), and the resistance to large shifts in behaviour captured by the inertia penalty \( \lambda_{\alpha} \cdot |I_B - I_A| \), where \( \lambda_{\alpha} \) is an agent-specific inertia coefficient:
\begin{align}
x_{AB}^{\alpha} = w_{\alpha} \cdot \tilde{S}_{AB}^{\alpha} + (1 - w_{\alpha}) \cdot A_{AB}^{\alpha} - \lambda_{\alpha} \cdot |I_B - I_A|.
\label{eq:x}
\end{align}
The parameter \( w_{\alpha} \) determines the relative weight assigned to social influence compared to attitudinal motivation, allowing agents to differ in how strongly they respond to peer behaviour versus internal environmental values. The term \( |I_B - I_A| \) quantifies the magnitude of the proposed intensity shift.
To ensure comparability with environmental attitude, which is bounded in the interval \([-1, 1]\), the social influence index \( S_{AB}^\alpha \) is clipped to the same range. Specifically, we define the adjusted social influence as:
\begin{equation}
\tilde{S}_{AB}^{\alpha} =  \min\left(1, \max\left(-1, 2 \cdot S_{AB}^{\alpha}\right)\right).
\end{equation}
This guarantees that both variables operate on a consistent scale, facilitating interpretation and integration within the model. All behavioural parameters including their value range are listed in Table \ref{tab:parameters}.
\begin{table}[ht!]
\centering
\begin{tabular}{p{4.5cm} p{8.7cm} p{1.2cm}}
\hline
\textbf{Parameter} & \textbf{Interpretation} & \textbf{Range} \\
\hline
Environmental attitude $A_{\alpha}$ 
& Land managers’ preference for extensification or intensification based on environmental values 
& $[-1,\,1]$ \\

Inertia coefficient $\lambda_{\alpha}$ 
& Strength of land managers’ resistance to changing land management intensity 
& $[0,\,1]$ \\

Importance of social norms $w_{\alpha}$ 
& Relative weight of social norms in decision-making compared to environmental attitude 
& $[0,\,1]$ \\

Critical mass $CM_{int,\alpha}$, $CM_{ext,\alpha}$ 
& Proportion of adopting neighbours required for a land manager to consider intensification or extensification, respectively 
& $[0,\,1]$ \\

Upper limit of the giving-in threshold $L_{\alpha}$ 
& Upper bound of the giving-in threshold. Lower values increase the relative importance of economic competitiveness in decision-making. 
& $[0,\,1]$ \\

Number of teleconnections $N_{tele}$ 
& Number of random long-range links in the social network in addition to spatial neighbourhood links 
& $\mathbb{N}_0$ \\

Neighbourhood radius $S_{nb}$ 
& Moore neighbourhood radius used to define local social interactions 
&$\mathbb{N}$ \\

Demand for material services $D_{mat}$ 
& Societal demand for material ES (e.g. crops, timber) 
& $\mathbb{N}_0$ \\

Demand for non-material services $D_{nm}$ 
& Societal demand for non-material ES (e.g. recreation, biodiversity) 
& $\mathbb{N}_0$ \\
\hline
\end{tabular}
\caption{Behavioural and external parameters of the model. Parameters with subscript $\alpha$ are agent-specific behavioural parameters that may vary across the landscape. All other parameters are global and exogenous to individual decision-making.}
\label{tab:parameters}
\end{table}

\subsubsection{Integration into an Agent-Based Land Use Model}

\begin{figure}[!t]
\centering
\includegraphics[width=1\linewidth]{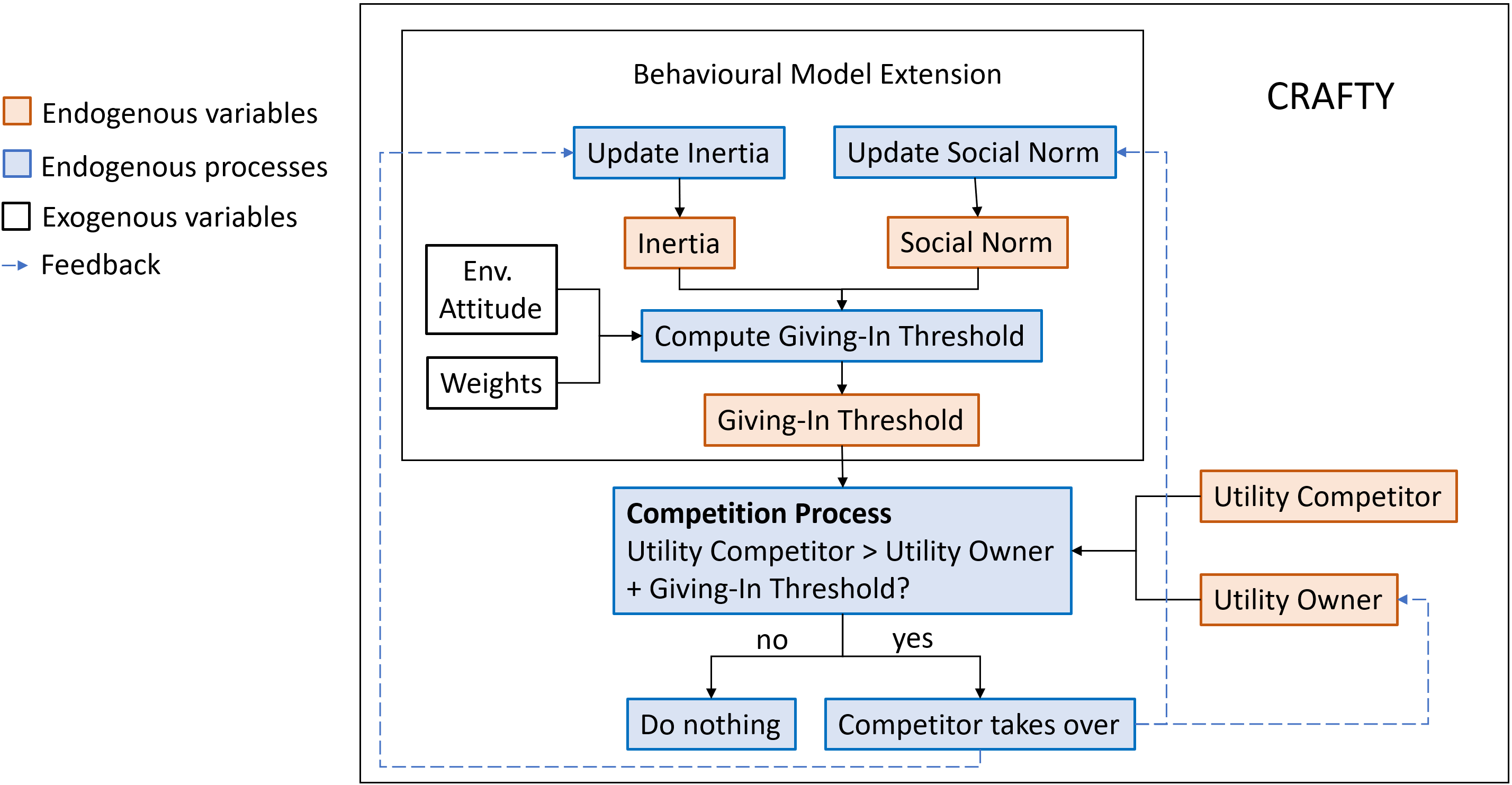}
\caption{Simplified architecture of the behavioural model extension integrated into CRAFTY. The behavioural weights include the importance of social norms, the inertia coefficient, and the upper limit of the giving-in threshold. Only those components of CRAFTY that are relevant for the behavioural extension are shown. A full sequence diagram of CRAFTY and a detailed flow diagram of the behavioural extension are provided in the Appendix (Figures~\ref{fig:flow_diagramm} and~\ref{fig:ext_CRAFTY_embedded}).}
\label{fig:behav_model_integration}
\end{figure}

To demonstrate the applicability of the behavioural model, we integrate it into the agent-based land use modelling framework CRAFTY (Competition for Resources between Agent Functional TYpes) \citep{murray-rust_open_2014, brown_agentbased_2022}. CRAFTY enables exploration of possible future land use scenarios, supports policymakers and researchers in assessing the effects of policies or market changes, and evaluates trade-offs between ES (ES) such as food production, carbon sequestration, and biodiversity conservation. CRAFTY is especially suited to apply behavioural submodels due to its partly modular structure and high flexibility. CRAFTY simulates land use dynamics across a spatial grid, where each cell is managed by an agent that produces ES using local biophysical and socio-economic resources, referred to as capitals. Agents belong to predefined Agent Functional Types (AFTs), which determine their production functions and behavioural traits. These AFTs can reflect different land use categories and intensities (e.g., intensive farming, conservation), and can be parameterised to express behavioural heterogeneity.

At each time step, agents compete for land based on their ability to meet service demands, illustrated by Figure \ref{fig:competition}. Land use change occurs when a competing agent’s utility $U_B^{\alpha}$ at a cell exceeds that of the incumbent $ U_A^{\alpha}$ by more than a giving-in threshold \( GIT_{AB}^{\alpha} \),
\begin{align}
U_B^{\alpha} - U_A^{\alpha} > GIT_{AB}^{\alpha}.
\end{align}
With the integration of our behavioural extension, this giving-in threshold is dynamically computed using the behavioural influence function introduced in the previous section, see Equation \ref{eq:GIT}. In CRAFTY, the behavioural attributes of land managers used within the model extension, indexed by $\alpha$, are stored as cell properties and can be parametrized specifically for each geographic location, while agents represent land management practices. 
Consequently, the current land management practice at a cell $\alpha$ is expressed by the incumbent agent's AFT denoted by $A$ and the prospective land management practise by the competitor's AFT, denoted by $B$.

While CRAFTY allows for changes between both land use intensities and categories such as afforestation and deforestation, the behavioural model focuses specifically on intensity transitions. Figure~\ref{fig:behav_model_integration} provides a simplified overview of how the behavioural model extension is integrated into CRAFTY. A sequence diagram of the full model version with integrated extension can be found in the Appendix, see Figure \ref{fig:ext_CRAFTY_embedded}.

\subsection{Experimental Setup}
For our experiments, we embed the behavioural extension in a light version of CRAFTY implemented in NetLogo \citep{wilensky_netlogo_1999} to facilitate the identification of causal model mechanisms. In this simplified version, capitals and demand levels remain static throughout the simulation, and land use change emerges solely through the competition process, omitting the giving-up and land allocation processes present in the full CRAFTY framework (Figure \ref{fig:ext_CRAFTY_embedded}).

We further restrict the analysis to transitions in land management intensity within a single land use category, such as the adoption of more extensive management practices by farmers or forestry practitioners.

\begin{figure}[!t]
\centering
\subfigure[Productive capital distribution]{
  \includegraphics[width=0.45\linewidth]{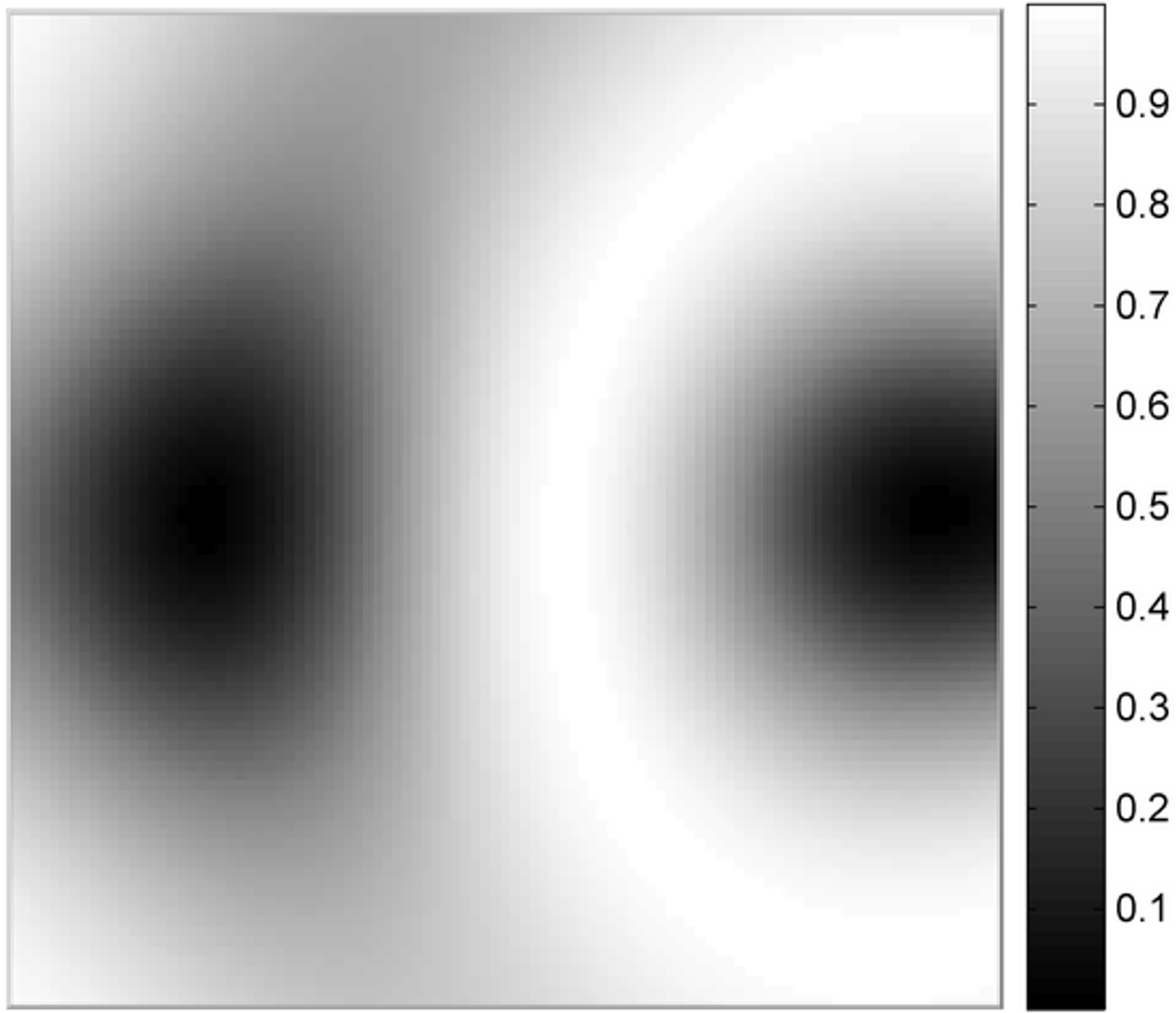}
  \label{fig:crop}
}
\hfill
\subfigure[Natural capital distribution]{
  \includegraphics[width=0.45\linewidth]{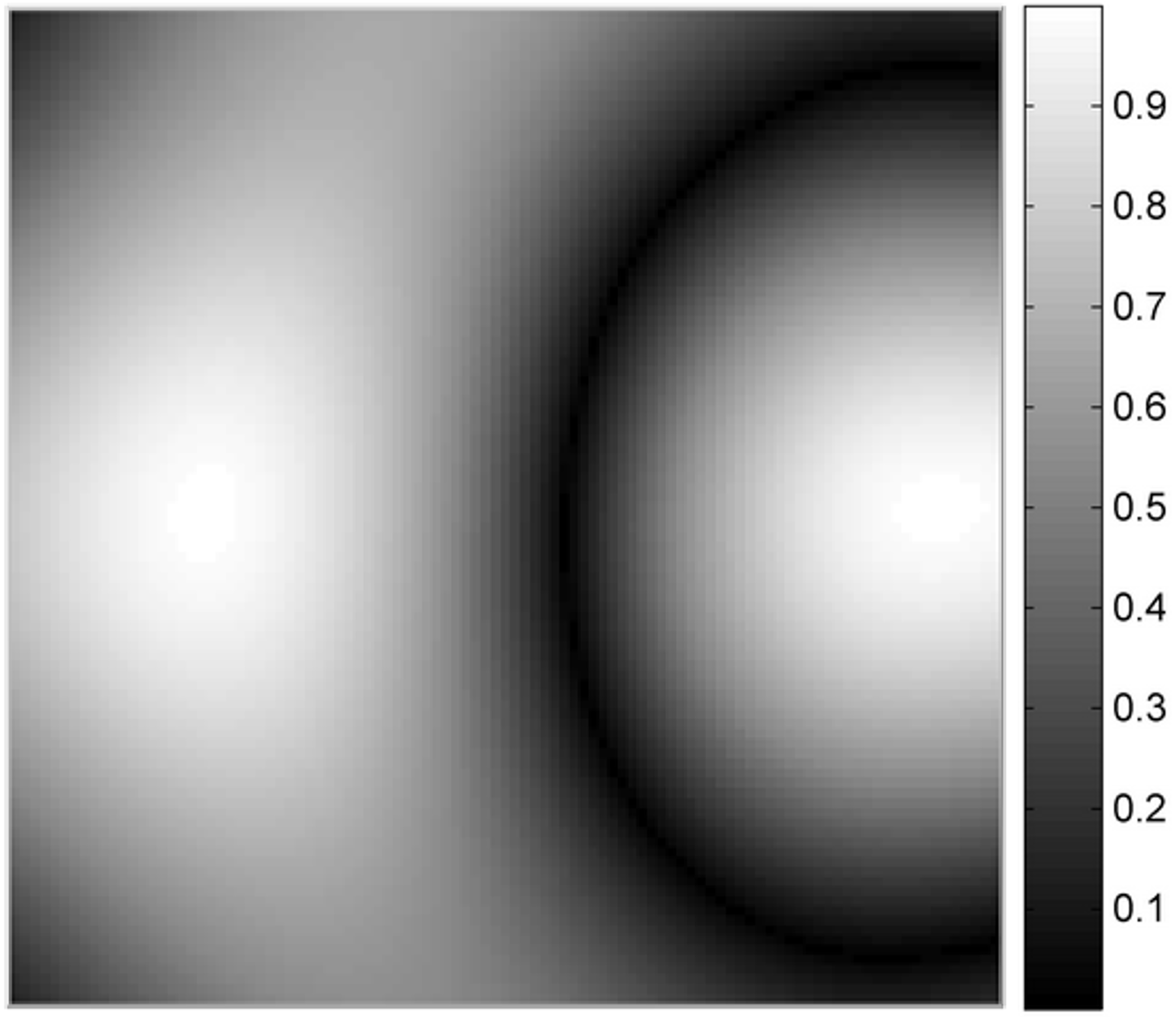}
  \label{fig:natural}
}
\caption{Capital distribution across the modelled landscape. Lighter colours indicate higher capital levels.}
\label{fig:capitals}
\end{figure}

To investigate how behavioural drivers shape land use outcomes, we use a stylized landscape configuration designed to minimise confounding influences. The simulated landscape is a 101×101 cell grid with spatially heterogeneous distributions of productive and natural capitals (Figure~\ref{fig:capitals}). Conceptually, the landscape can be thought of as a region with two mountains rich in natural capital, providing non-material ES such as recreation, surrounded by a valley with higher productive capital that favours more intensive land use.

Each cell is initially managed by a single land manager, randomly assigned to one of three Agent Functional Types (AFTs): high intensity managers, medium intensity managers, or conservationists. These AFTs vary in their sensitivity to different forms of capital and in the types and levels of ES they produce (Table~\ref{tab:AFTs}).
Agents compete for land based on their ability to produce ES, which represent material services (e.g., food, timber) and non-material services (e.g., cultural services). In our model runs, we simulate a single sector system involving only one land use category. This can represent for example forestry practitioners producing timber or farmers producing crops using different intensity levels of land management, for instance by varying their fertilizer use. For simplicity, production is determined by linear functions of local capitals:
\begin{equation}
P_{mat, A}^{\alpha} = s_{prod}^A \cdot C_{prod}^{\alpha}, \quad P_{nm, A}^{\alpha} = s_{nat}^A \cdot C_{nat}^{\alpha},
\end{equation}
where \(P_{mat, A}^{\alpha}\) and \(P_{nm, A}^{\alpha}\) are the material and non-material ES produced by agent \(A\) on cell \(\alpha\), \(C_{prod}^{\alpha}\) and \(C_{nat}^{\alpha}\) are the productive and natural capital levels at that cell, and \(s_{prod}^A\), \(s_{nat}^A\) are AFT-specific sensitivities (Table~\ref{tab:AFTs}). 
\begin{table}[ht!]
\centering
\caption{Sensitivity of agent functional types to productive and natural capital.}
\label{tab:AFTs}
\begin{tabular}{lcc}
\hline
\textbf{Agent Functional Type} & \textbf{Productive Capital} & \textbf{Natural Capital} \\
\hline
High Intensity      & 1.0 & 0.0 \\
Medium Intensity   & 0.5 & 0.5 \\
Conservation       & 0.0 & 1.0 \\
\hline
\end{tabular}
\end{table}
All behavioural mechanisms, production functions, and the demand-driven competition framework are implemented symmetrically with respect to conservation and high intensity land use. That is, conservation is modelled as the mirror process of high intensity land use: both respond to demand for ES and are influenced by environmental attitudes in structurally analogous ways.

In each simulation step, 5\% of the cells are randomly selected for potential land use change as discussed in \citep{brown_societal_2019}. The managing agent at each selected cell evaluates whether to intensify, extensify, or maintain the current land management intensity. Decisions are guided by the competition process with integrated behavioural model extensions as described in the Methods Section. Simulations run until the system reaches a steady or quasi-steady state.

\subsection{Evaluation Methods}
To assess the influence of socio-psychological drivers on land management intensity shares, ES provision, and spatial patterns, we combine a global Sobol sensitivity analysis with structured parameter sweeps and experiments in which environmental attitudes vary dynamically over time.

The Sobol sensitivity analysis \citep{sobol_global_2001} systematically varies behavioural parameters, social network configurations, and ES demand levels within predefined ranges (Table \ref{tab:sobol_ranges}), which together influence competition among land managers. We compute first-order, second-order, and total-effect Sobol indices to identify which parameters most strongly shape system-level outcomes and how interactions between parameters contribute to these outcomes. As outcome metrics, we consider the relative share of each land management intensity as well as the total supply of material and non-material ES.
The upper limit of the giving-in threshold is fixed at $L=1$ in the sensitivity analysis. This choice ensures that the behavioural components introduced through the giving-in threshold operate on the same value range $[0,1]$ as the economic competition term, expressed by the utility difference between competitor and owner $(U_B - U_A)$. Because the focus of this study lies on understanding the role and interaction of behavioural drivers, fixing $L$ at its maximum avoids artificially constraining behavioural influence relative to economic incentives.
All sensitivity analyses are conducted using the Python library SALib \citep{herman_salib_2017}.

In addition to the global sensitivity analysis, we conduct structured parameter sweep experiments to explore nonlinear responses, regime shifts, and interaction effects that are difficult to capture with variance-based metrics alone. These sweeps vary selected behavioural parameters—most notably environmental attitude and the weight of social norms—while holding other parameters constant, allowing us to map distinct land use regimes and identify narrow transition zones between them. Outcome metrics are evaluated after the system has reached a stable configuration, ensuring that reported patterns reflect long-term dynamics rather than transient behaviour. We analyse compositional and spatial outcomes, including the final shares of each land management intensity and intra-patch connectivity, measured using the MESH-CWA index \citep{justeau-allaire_refining_2024}. 

To analyse path dependence and hysteresis, we run further experiments in which the environmental attitude changes dynamically over time. In these simulations, attitude is gradually increased and subsequently decreased, enabling us to assess whether the system returns to its original state or remains locked into an alternative configuration. 

\section{Results}

\subsection{Asymmetries from Landscape–Behaviour Feedbacks}
\begin{figure}[!t]
\centering
\subfigure[Conservation]{
  \includegraphics[height=3.8cm]{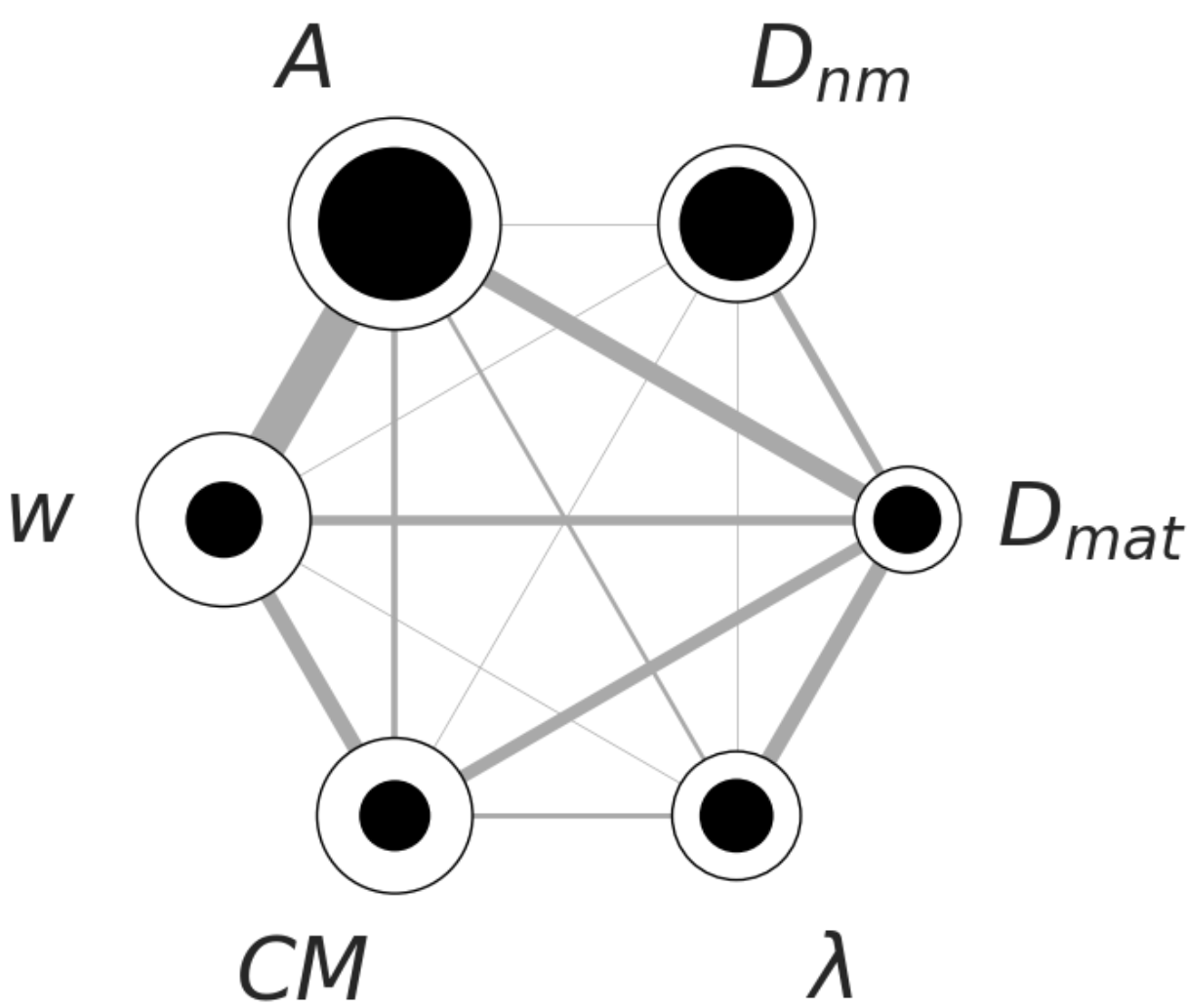}
  \label{fig:LI_circle}
}
\hfill
\subfigure[Medium intensity]{
  \includegraphics[height=3.8cm]{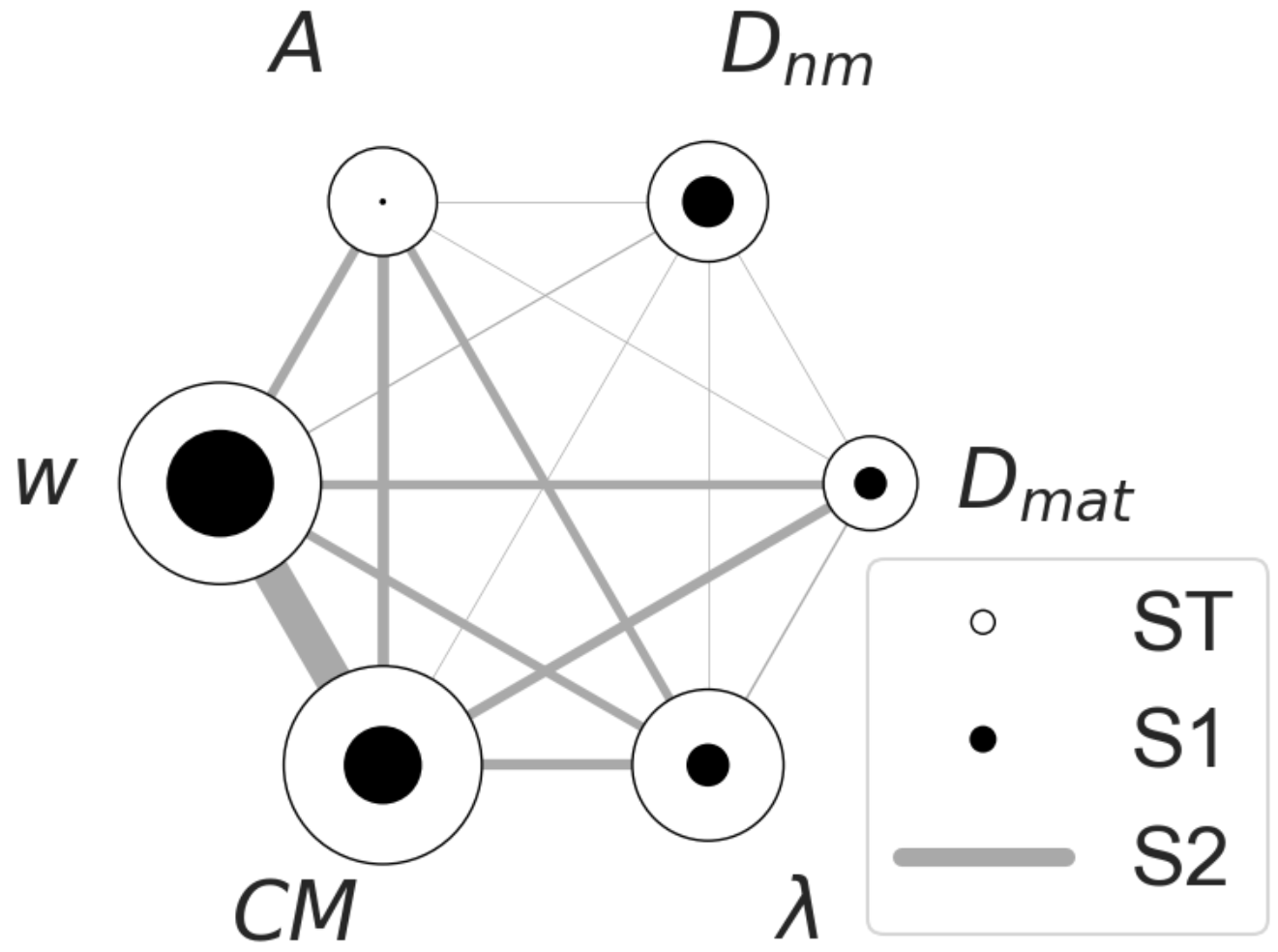}
  \label{fig:MI_circle}
}
\hfill
\subfigure[High intensity]{
  \includegraphics[height=3.8cm]{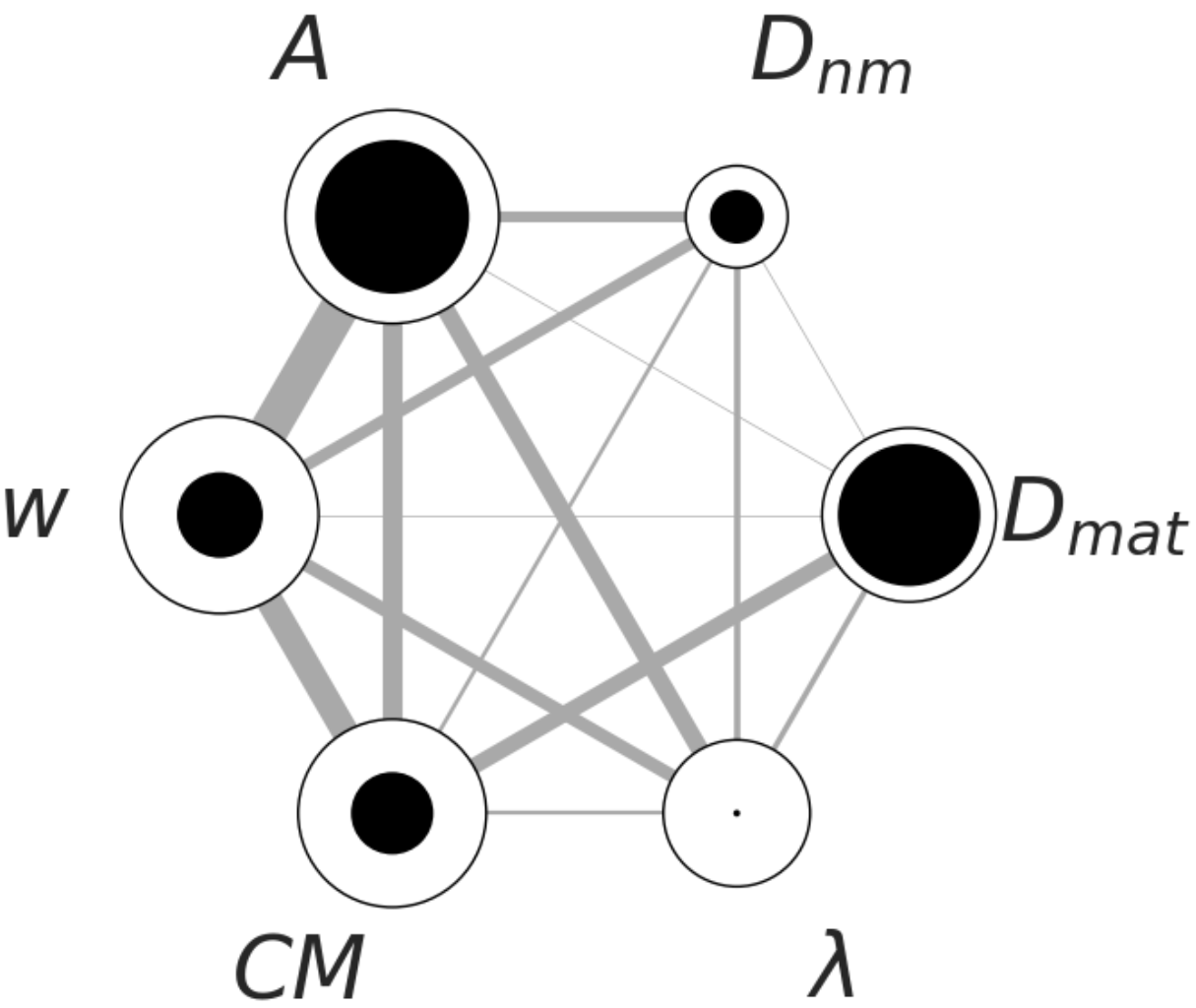}
  \label{fig:HI_circle}
}
\caption{First-order (S1), second-order (S2), and total-effect (ST) sensitivity indices from a global sensitivity analysis of intensity shares, varying behavioural parameters and demand levels (Table~\ref{tab:sobol_ranges}). According to the SA, intensity shares are independent of initial land use distribution, steepness of the logistic function $k$ and network characteristics (number of teleconnections $N_{tele}$ and neighbourhood radius $S_{nb}$). The upper limit of the giving-in threshold is set to $L=1$ for all cells. Node and edge sizes are proportional to the corresponding Sobol sensitivity indices, with larger circles or wider lines indicating greater influence on the outcome.}
\label{fig:three_intensity_circles}
\end{figure}

\begin{figure}[!t]
\centering
\subfigure[Non-material]{
  \includegraphics[height=5.3cm]{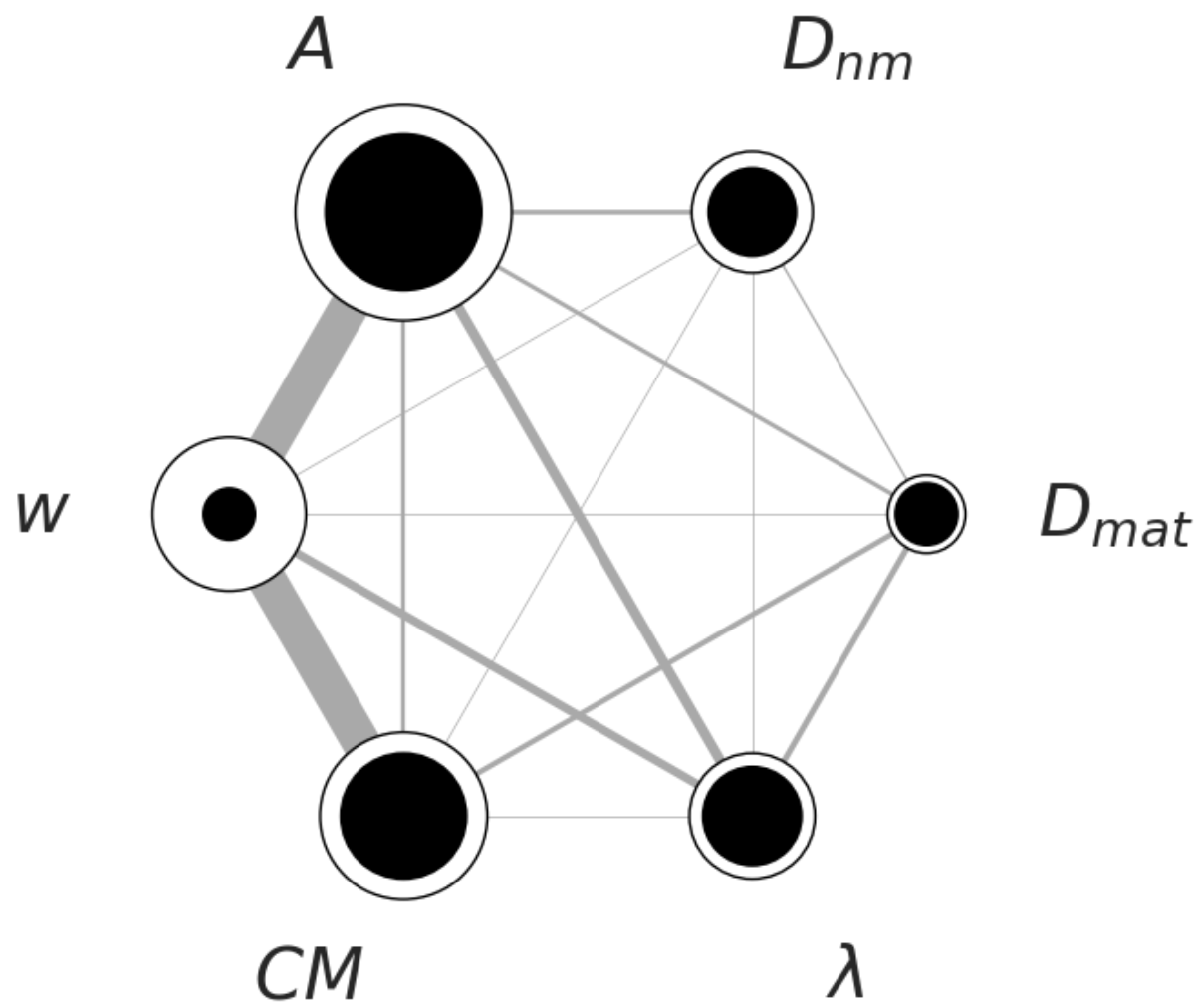}
  \label{fig:nm_circle}
}
\hfill
\subfigure[Material]{
  \includegraphics[height=5.3cm]{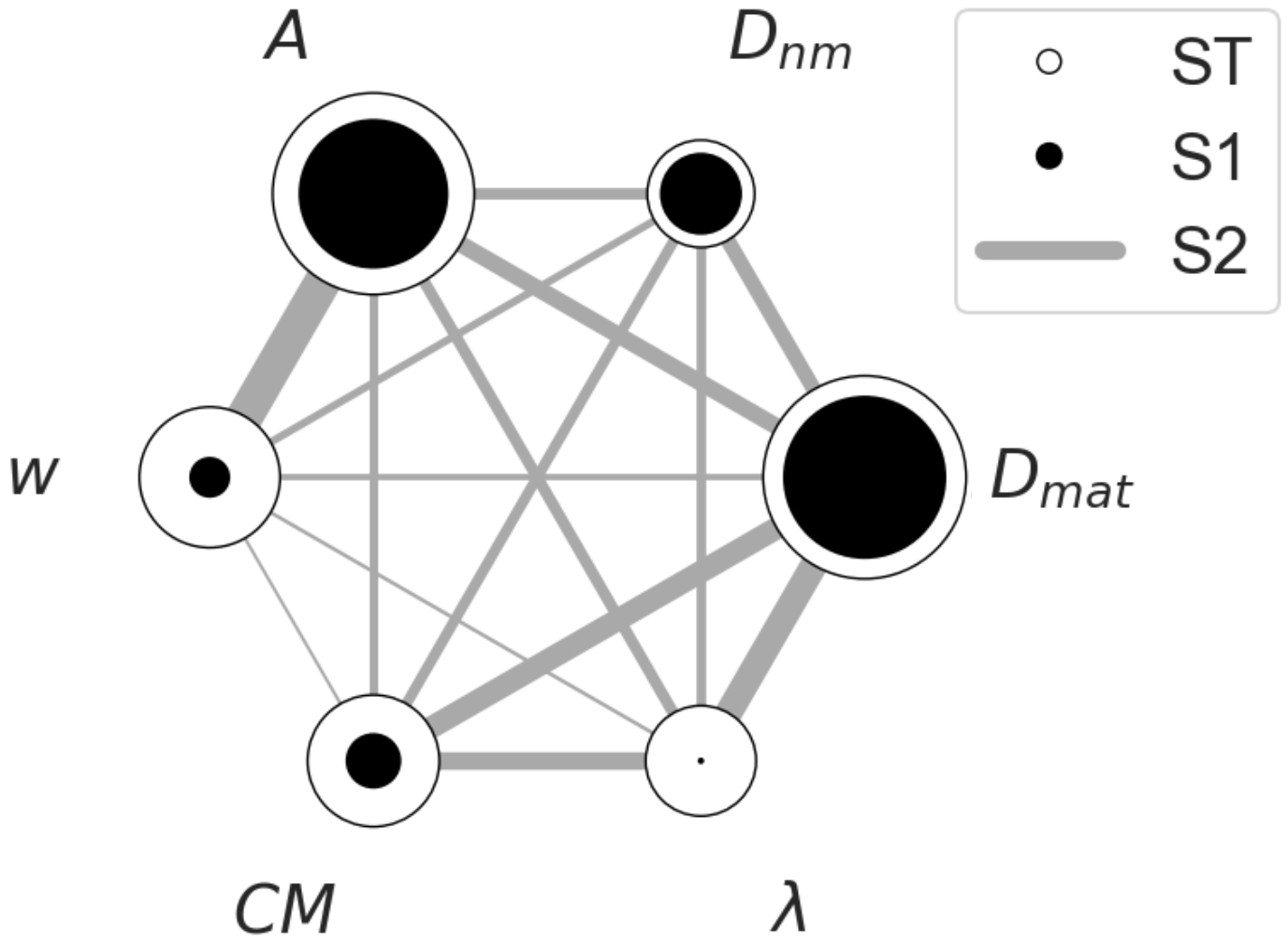}
  \label{fig:mat_circle}
}
\caption{First-order (S1), second-order (S2), and total-effect (ST) sensitivity indices from a global sensitivity analysis of total supply of non-material and material ES, varying behavioural parameters and demand levels (Table~\ref{tab:sobol_ranges}). According to the sensitivity analysis, total ES supply is independent of the initial land use distribution, the steepness of the logistic function $k$, and network characteristics (number of teleconnections $N_{tele}$ and neighbourhood radius $S_{nb}$). The upper limit of the giving-in threshold is set to $L=1$ for all cells. Node and edge sizes are proportional to the corresponding Sobol sensitivity indices, with larger circles or wider lines indicating greater influence on the outcome.}
\label{fig:two_ES_circles}
\end{figure}

 Although the model mechanisms are formally symmetric, conservation being driven by non-material ES demand and positive environmental attitudes, and high intensity land use by material demand and negative attitudes, an asymmetry emerges in the system’s sensitivity to input parameters. The share of high intensity land use responds more strongly to changes in material demand than the conservation share does to non-material demand (Figure \ref{fig:three_intensity_circles}). This pattern becomes even clearer when looking at total ES provision: material services closely follow demand levels, whereas non-material services are less sensitive to their demand and are instead more strongly influenced by environmental attitudes (Figure \ref{fig:two_ES_circles}). The reason for this asymmetry lies in the spatial distribution of capitals (Figure~\ref{fig:capitals}) as all other system parameters are varied symmetrically. Since productive capital, which is essential for high management intensity land use, is more widely available in the model landscape, high intensity managers predominantly occupy land rich in productive capital and are therefore highly responsive to economic signals such as material demand. Taken together, this finding highlights the nonlinear interactions between spatial resource configurations, agent behaviour, and external demand, showing that even when mechanisms are designed to be symmetric, emergent land use patterns can diverge substantially due to landscape heterogeneity and behavioural feedbacks.

\subsection{Interactions between Behavioural Drivers}

Surprisingly, a higher environmental attitude does not consistently lead to more conservation-oriented land use, despite the direct model rule that lowers the giving-in threshold for extensification under positive attitudes. Likewise, decreasing attitudes do not necessarily result in a higher share of high intensity land use.

This non-monotonic relationship arises from complex interactions between environmental attitude and other behavioural drivers— most notably social norms. The Sobol analysis identifies environmental attitude and the weight of social norms as the two parameters with the strongest second-order interaction effects on the conservation and high intensity shares revealing the importance of their interactions in shaping these outcomes (Figure \ref{fig:three_intensity_circles}).

\begin{figure}[!t]
    \centering
    \includegraphics[width=1\linewidth]{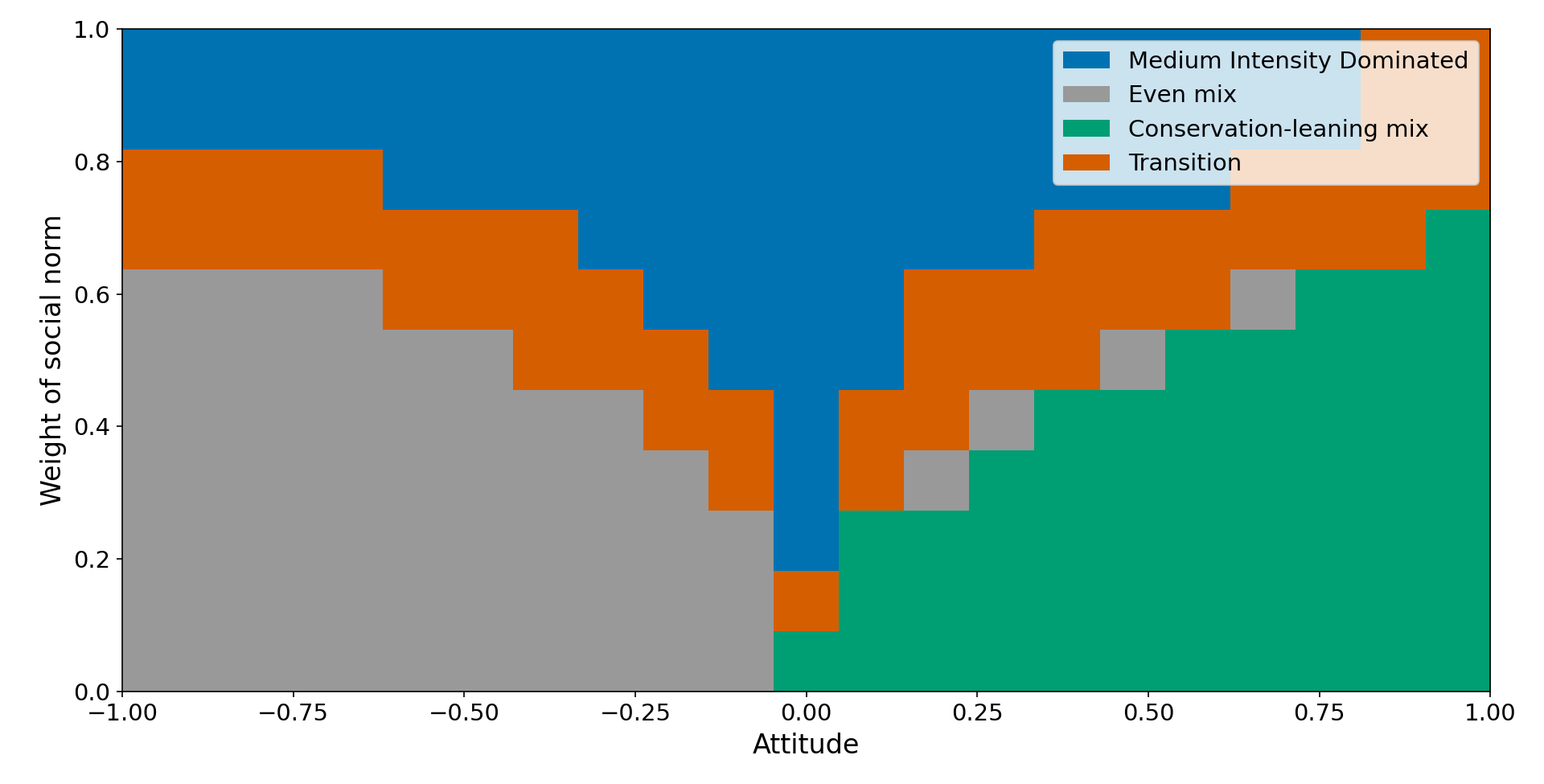}
    \caption{Regime map of dominant land management intensity compositions after stabilization as a function of mean environmental attitude and weight of social norm. 
    Blue indicates a medium intensity dominated regime (Medium Intensity Share $\approx$ 0.8, Conservation Share $\approx$ 0.1, High Intensity Share $\approx$ 0.0), 
    grey an even-mix regime (Conservation Share $\approx$ Medium Intensity Share $\approx$ High Intensity Share $\approx$ 0.33), 
    green a conservation-leaning regime (Conservation Share $\approx$ 0.4, Medium Intensity Share $\approx$ 0.3, High Intensity Share $\approx$ 0.3), 
    and orange a transition regime (Medium Intensity Share $\approx$ 0.4--0.7, Conservation Share $\approx$ 0.2--0.3, High Intensity Share $\approx$ 0.1--0.2). Simulation parameters as listed in Table \ref{tab:parameters_all}.}
    \label{fig:regimes}
\end{figure}

The regime plot (Figure \ref{fig:regimes}) illustrates these interactions. When environmental attitude increases from negative to moderate values, the share of conservation declines while medium intensity land use expands, particularly at intermediate to high values of $w$. This pattern can be explained by the role of social norms, which tend to promote medium intensity land use as a compromise between the extremes of conservation and high intensity. Their influence is strongest when attitudes are moderate—when agents are more receptive to peer influence—leading to local convergence around medium intensity practices and a temporary reduction of both extreme land use types.

The second mechanism involves demand-driven competitiveness. At high attitude values, conservation expands due to lower thresholds for extensification, but this expansion creates an undersupply of material ES, prompting compensatory growth in high intensity land use. A mirrored dynamic occurs for negative attitudes, where the expansion of high intensity land use reduces non-material services and subsequently increases conservation.

\begin{figure}[!t]
    \centering
    \includegraphics[width=1\linewidth]{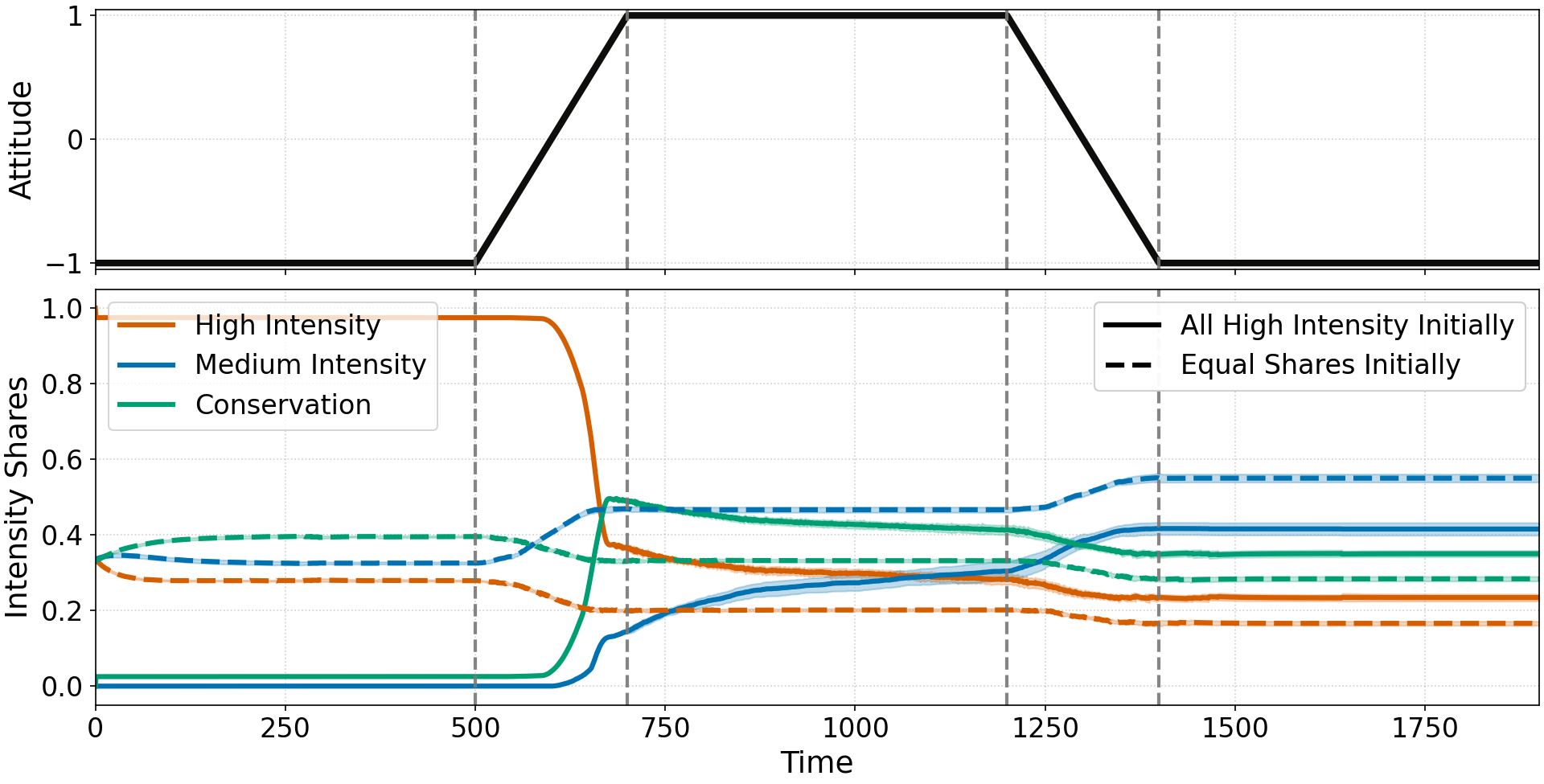}
    \caption{Temporal evolution of environmental attitude (top) and relative intensity shares (bottom) under two different initial land use distributions. Shaded areas indicate one standard deviation around the mean. Vertical dashed lines mark attitude transition phases shared across both panels. Simulation parameters as listed in Table \ref{tab:parameters_all}.    }
    \label{fig:dyn_att}
\end{figure}

Simulations with dynamically changing attitudes show similar nonlinear feedbacks over time (Figure \ref{fig:dyn_att}).
When environmental attitude declines, one might expect an increase in high intensity land use. Instead, the shift toward more productive preferences often triggers a transition from conservation to medium intensity practices rather than to high intensity.
This occurs because the resulting undersupply of non-material ES favours the spread of medium intensity land use, which balances material production and non-material service provision.
Through social influence, these medium intensity practices then spread across the landscape, further reducing the share of high intensity management. 

Although stronger social influence often reduces the share of strict conservation, the supply of non-material ES can still increase slightly (Figure~\ref{fig:SA_dependencies}).
This seemingly counterintuitive effect arises because medium intensity land managers, who expand under stronger normative influence, contribute to both material and non-material services, partially offsetting the decline in conservation.

Overall, these results demonstrate that the emergent land management dynamics cannot be understood from the direct, rule-based link between attitude and decision thresholds alone. Instead, they result from higher-order feedbacks between attitude, social norms, and economic competitiveness that jointly shape the collective behavioural outcomes.

While these mechanisms operate at the level of behavioural feedbacks, social norms also have a distinctly spatial dimension: they shape how behaviours diffuse across the landscape and how clusters of similar management practices form and persist.
In the following, we examine how the strength and spatial reach of social influence affect both the composition and the spatial configuration of land use intensities.

\begin{figure}[!t]
    \centering
    \includegraphics[width=1\linewidth]{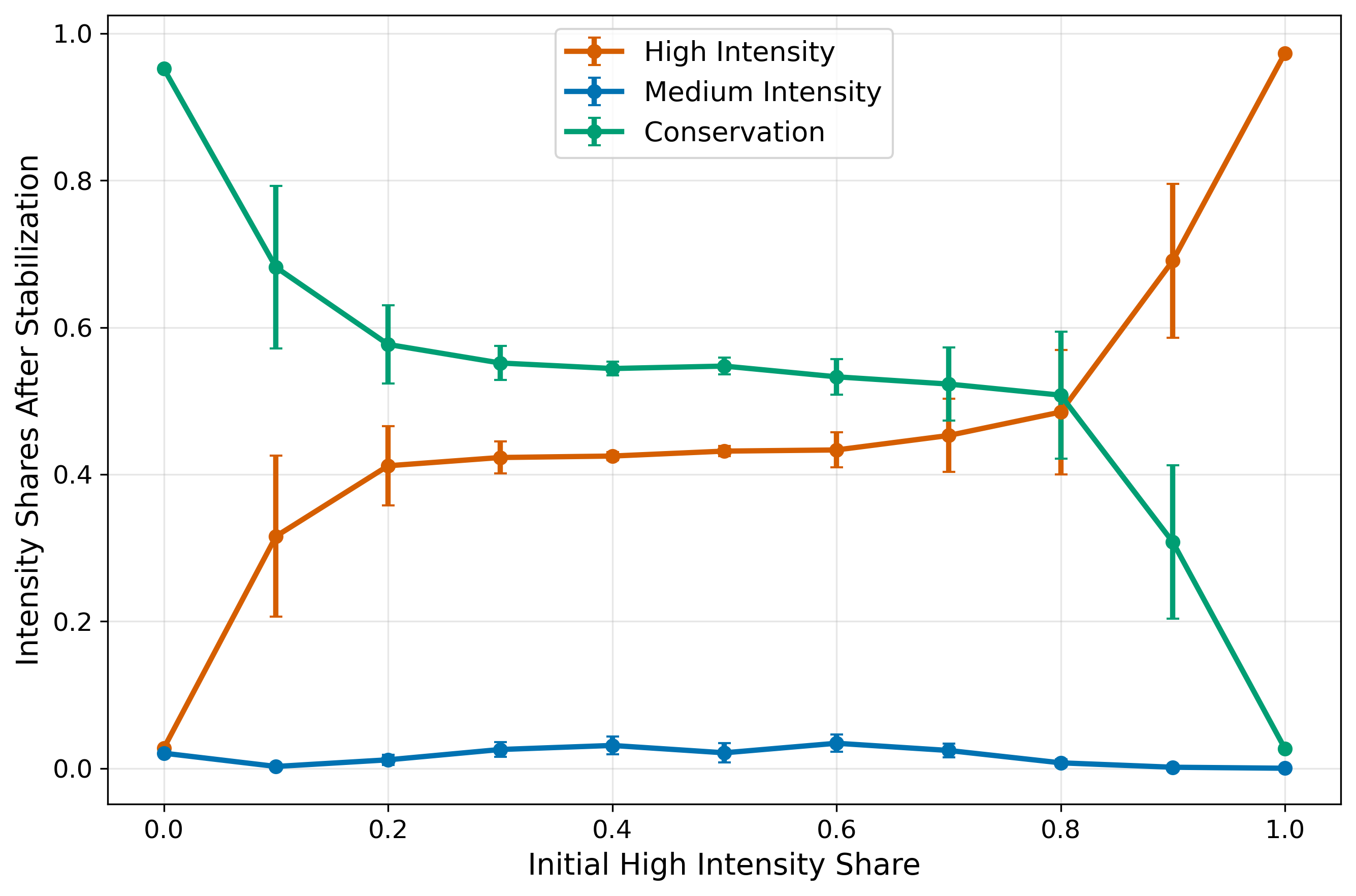}
    \caption{Intensity shares after stabilization as a function of the initial high intensity share. 
The initial medium intensity share is set to zero in all runs. 
Error bars represent one standard deviation across replications.  
A full analysis including variation of the initial medium intensity share is provided in the ternary plot in Appendix Figure ~\ref{fig:ternary_final_shares}. 
Simulation parameters as listed in Table~\ref{tab:parameters_all}.}
    \label{fig:initial_conditions}
\end{figure}

\subsection{Social Norms and Network Characteristics}
Social norms tend to promote medium intensity land use as a socially acceptable compromise between the two extremes of conservation and high intensity management.
However, this stabilising influence requires that medium intensity practices already exist in the landscape or are encouraged by other factors.
When the initial conditions contain only the two extremes — high intensity and conservation — medium intensity land management rarely emerges, even when the initial distribution of the two extremes is balanced (Figure \ref{fig:initial_conditions}).
Once a sufficient number of medium intensity adopters is present, however, it spreads much more readily than the other two practices.
The ternary plot of final intensity shares (Figure \ref{fig:ternary_final_shares}) illustrates this asymmetry: a large proportion of initial-condition combinations lead to landscapes dominated by medium intensity, whereas configurations dominated by high intensity or conservation practices are far less common.
This finding suggests that medium intensity land use represents a particularly robust social equilibrium — one that can diffuse rapidly once a critical mass of adopters is reached.

\begin{figure}[!t]
\centering
\subfigure[$S_{nb}=1$]{
  \includegraphics[width=0.17\linewidth]{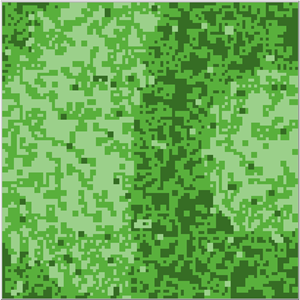}
  \label{fig:rad1_bal}
}
\hfill
\subfigure[$S_{nb}=2$]{
  \includegraphics[width=0.17\linewidth]{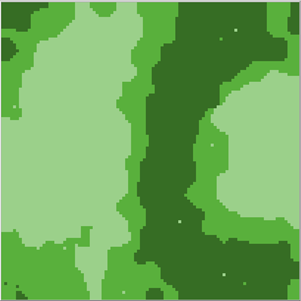}
  \label{fig:rad2_bal}
}
\hfill
\subfigure[$S_{nb}=3$]{
  \includegraphics[width=0.17\linewidth]{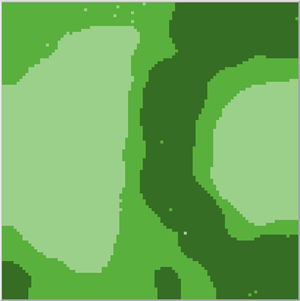}
  \label{fig:rad3_bal}
}
\hfill
\subfigure[No norms]{
  \includegraphics[width=0.17\linewidth]{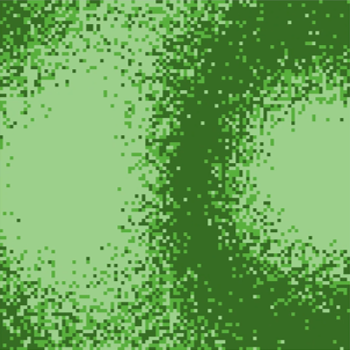}
  \label{fig:no_norms}
}
\hfill
\subfigure[$N_{tele}=4000$]{
  \includegraphics[width=0.17\linewidth]{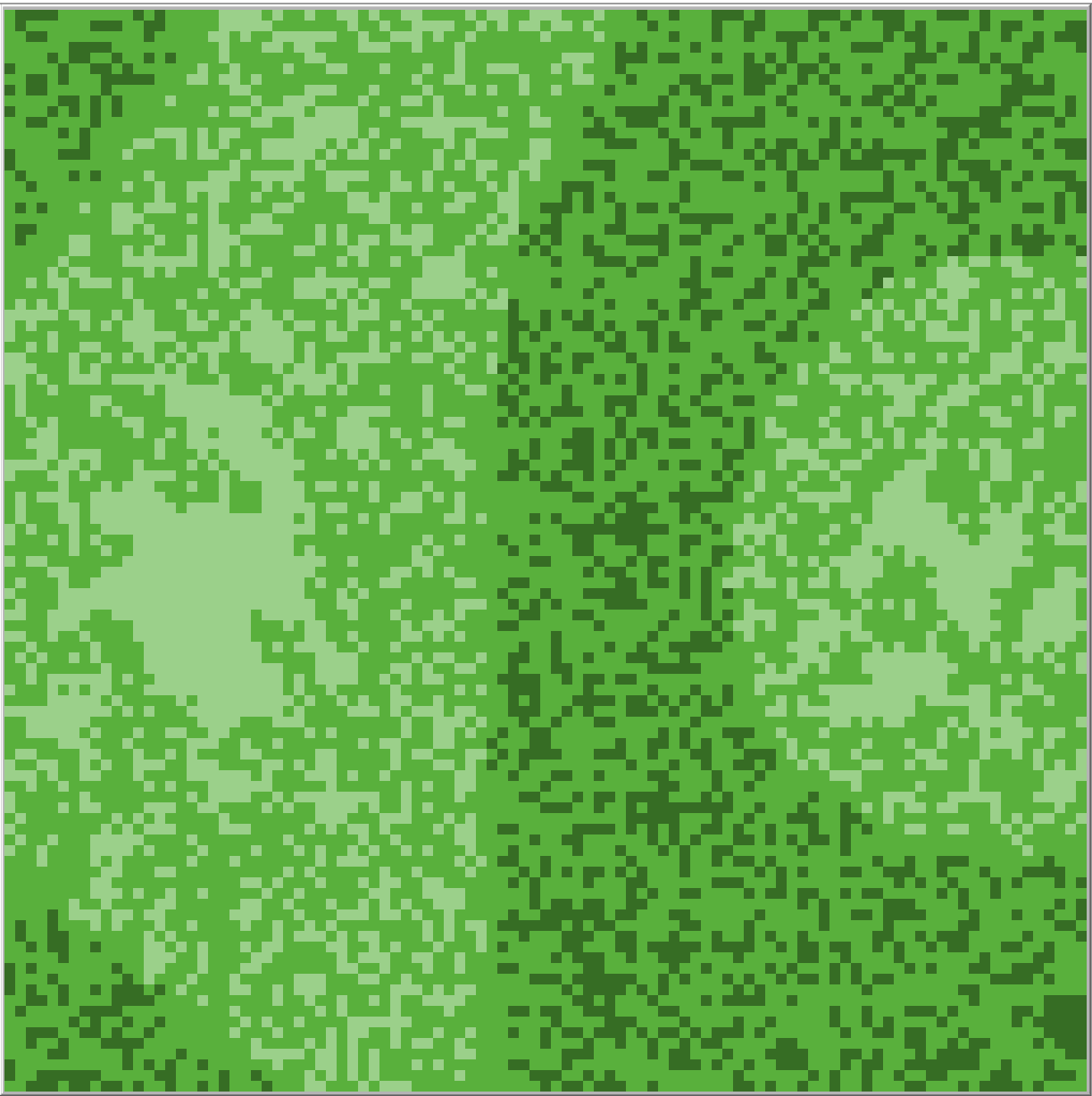}
  \label{fig:tele}
}

\caption{Land use patterns after stabilization for varying neighbourhood sizes ($S_{nb}$), including a benchmark case without social norms and a teleconnected case ($S_{nb}=2$, $N_{tele}=4000$, corresponding to roughly 40\% of the number of patches). All other cases have no teleconnections. Simulation parameters are listed in Table~\ref{tab:parameters_all}.}
\label{fig:balanced_maps}
\end{figure}

Social influence not only alters the overall distribution of land use types but also generates distinct spatial patterns within the landscape. In the absence of social norms, land use intensities are primarily structured by the underlying capital distribution: high intensity land use concentrates in areas with abundant productive capital, conservation dominates regions rich in natural capital, and transitional zones exhibit highly variable and unstable patterns (Figure~\ref{fig:no_norms}). When social norms are introduced, however, new spatial configurations emerge. While the overall proportion of each land management intensity remains largely unaffected by neighbourhood size or the number of teleconnections (Sobol Sensitivity Analysis Figure \ref{fig:three_intensity_circles}), the spatial arrangement of those intensities changes substantially. Clusters of medium intensity land use appear throughout the landscape when neighbourhood influence operates at a small spatial radius (Figure \ref{fig:rad1_bal}). As the neighbourhood size increases, these scattered clusters coalesce into large, continuous patches of medium intensity land use, forming cohesive bands in the transitional zones between high- and low-intensity areas (Figure \ref{fig:rad2_bal} and \ref{fig:rad3_bal}). In contrast, introducing teleconnections — long-range social links between distant agents — reduces local spatial coherence and mixes behavioural influence across the entire landscape (Figure \ref{fig:tele}).
As a result, the previously distinct clusters of medium intensity land use dissolve into a more fine-grained mosaic, with scattered medium intensity patches overlaying the sharply defined high- and low-intensity zones.
This pattern indicates that global social connectivity promotes the diffusion of compromise behaviours at the expense of cohesive regional clustering.

These qualitative differences in spatial configuration are also captured quantitatively by intra-patch connectivity (Figure~\ref{fig:connectivity_balanced}). When social influence is weak and decision-making is dominated by economic competitiveness, expressed by a low upper limit of the giving-in threshold $L$ under a fixed social norm weight ($w=1$), connectivity remains low. This reflects a fragmented landscape structure in the transitional zones between areas dominated by productive and natural capital (Figure~\ref{fig:no_norms}). As the relative influence of social norms increases, connectivity rises markedly, indicating the formation of more cohesive land use patches, supporting the visual observations from the maps (Figure \ref{fig:rad3_bal}). However, beyond a certain threshold, further increases in $L$ lead to a gradual decline in connectivity, as high conformity requirements suppress behavioural change. Nevertheless, connectivity remains substantially higher than in the absence of social norms, underscoring their critical role in shaping spatial structure. Across most values of $L$, connectivity is consistently higher when agents interact within a larger neighbourhood (size 4), indicating that broader social influence promotes stronger spatial coordination and the formation of more extensive clusters.

\subsection{Critical Transitions and Path Dependence}
When all land managers initially engage in the same land management intensity and most do not hold attitudes opposing this practice, the system can stabilise near its initial configuration, with production decoupled from demand (Figure \ref{fig:initial_conditions} and \ref{fig:dyn_att}).

In an experiment where environmental attitude is increased linearly over time, the system transitions from this decoupled regime to one in which supply follows demand once the prevailing environmental attitude exceeds a threshold that clearly opposes the dominant management practice (Figure \ref{fig:dyn_att}). Starting from a stable plateau dominated by high intensity land management, the share of this land management practice declines abruptly over a narrow range of attitude values, producing an S-shaped response.

The location and sharpness of this transition are confirmed by an additional parameter sweep varying both environmental attitude and the upper limit of the giving-in threshold $L$ (Figure \ref{fig:hif_heatmap_att_L}). The resulting plot reveals two distinct stable regimes separated by a well-defined transition in parameter space. For the value $L = 0.65$ used in the dynamic-attitude experiment, this boundary occurs at approximately $A \approx 0.2$, consistent with the critical transition zone observed before.

Extending this analysis, a two-dimensional exploration of environmental attitude and the weight of social norms reveals three distinct stable regimes of final land management shares — a conservation-leaning mix, a medium-intensity-dominated regime, and an even mix — separated by narrow transition zones in parameter space (Figure \ref{fig:regimes}). Within each regime, outcomes are relatively insensitive to small parameter changes, whereas transitions between regimes are abrupt, supporting the interpretation of the system as a multi-stable landscape structured by sharp regime boundaries.

The dynamic-attitude experiment further reveals a clear hysteresis effect: the system does not generally return to its initial state once attitudes are restored (Figure \ref{fig:dyn_att}). This asymmetry indicates path dependence — past configurations influence the attainable equilibria, and identical parameter settings can lead to qualitatively different outcomes depending on the system’s history. The discrepancy between the upward and downward trajectories becomes more pronounced under extreme initial conditions, but it remains visible even when the initial land use composition is balanced. These results highlight that transitions between behavioural regimes are not fully reversible: once a configuration of social norms and land use intensities becomes established, it tends to persist even after the underlying drivers change back.

Together, the abrupt attitude-driven transition between two stable regimes, resistance to reversal, and persistence of alternative land management configurations indicate the presence of lock-in effects. Reinforcing interactions between social norms, spatial clustering, and demand-driven competition stabilise prevailing land management patterns and limit the system’s responsiveness to changing attitudes once a new configuration becomes established.

Further evidence for path dependence is provided by the system’s sensitivity to initial land management composition. When a single management practice initially dominates the landscape, the system stabilises near its starting configuration (Figure \ref{fig:initial_conditions}), whereas introducing a minority practice can trigger disproportionate changes in long-term outcomes as it spreads across the landscape. Once the initial share of the minority practice exceeds a critical mass, the system converges to a characteristic stable configuration that is largely independent of initial conditions, indicating the presence of a basin of attraction. Outside this range, small changes in initial composition lead to large differences in final land management patterns.

The strength of this path dependence depends on the critical mass threshold governing social influence (Figure \ref{fig:critical_mass}). At high critical mass values, agents require strong neighbourhood consensus before changing practices, resulting in limited behavioural change and increased dependence on initial conditions. In contrast, at low critical mass values, behavioural change becomes socially unconstrained, and land management shares converge toward those of the purely economic baseline model without social norms. 

\section{Discussion}
Land managers’ behaviour is shaped by more than economic and structural factors alone, yet socio-psychological drivers remain underrepresented in many land use models. In this study, we introduced a behavioural model extension that incorporates empirically grounded mechanisms — environmental attitudes, descriptive social norms, and behavioural inertia — and examined their effects in a stylised land use setting.

The discussion proceeds in three steps. First, we synthesise the main emergent dynamics produced by the interaction of these socio-psychological drivers with economic competition and landscape structure. We then reflect on key modelling choices and the implications of the behavioural assumptions underlying the extension. Finally, we discuss limitations of the current approach and outline directions for future research.

\subsection{Emergent Dynamics}
\subsubsection{Asymmetries from Landscape–Behaviour Feedbacks}
The observed asymmetry between conservation and intensification responses to symmetrically varied behavioural and external input parameters highlights the importance of spatial heterogeneity in resource configurations for shaping aggregate land use dynamics. This finding shows that even simple models with symmetric behavioural processes can generate emergent outcomes that are strongly biased by the underlying resource landscape and by nonlinear feedbacks between demand, capital availability, and socio-psychological factors.
Furthermore, it reveals that in regions with a structural advantage of high intensity land management over conservation, interventions targeting non-material ES may require different incentives or disproportionately stronger signals to counterbalance this production-oriented advantage. While economic incentives are highly effective in increasing material ES provision, the supply of non-material ES is primarily driven by environmental attitudes and is therefore less responsive to purely economic drivers.

\subsubsection{Interactions between Behavioural Drivers}

Although the model directly links environmental attitudes to a lower giving-in threshold for extensification, the results show that pro-environmental attitudes do not necessarily translate into more conservation-oriented land management practices and can, in some cases, even reduce it. This aligns with empirical observations that positive attitudes alone cannot fully explain land managers’ adoption behaviour \citep{swart_meta-analyses_2023, diana_feliciano_decision_2025}. While a general positive correlation is often observed empirically, attitudes and intentions do not necessarily lead to behavioural change — reflecting the well-documented intention–behaviour gap in land use \citep{swart_meta-analyses_2023}. These findings highlight the importance of accounting for multiple motivational factors and their interactions when analysing behavioural outcomes.

\subsubsection{Social Norms and Network Characteristics}
Under strong social influence, medium intensity land use increases particularly in the transitional zones between regions dominated by high intensity and conservation practices. This suggests that agents converge toward socially acceptable compromise behaviours rather than adopting either extreme.

Descriptive social norms also lead to higher local coherence of land use intensities, producing spatial patterns comparable to empirical observations of clustered management practices in rural landscapes \citep{schmidtner_spatial_2012, lapple_spatial_2015}. The spatial agglomeration of land management intensities observed in the model arises from two interacting mechanisms: the adoption of land management practices follows the distribution of productive and natural capital, and the influence of descriptive social norms among neighbouring agents. The first determines the biophysical potential for land management intensification or conservation, producing clusters that follow gradients of soil fertility and natural suitability. The second reinforces these patterns through social contagion, as agents align their behaviour with the prevailing practices in their vicinity. This dual mechanism mirrors the empirical findings of Schmidtner et al. \citep{schmidtner_spatial_2012}, who explained the spatial distribution of organic farming in Germany by both social connectivity effects (peer influence) and biophysical connectivity effects (land suitability and climate).

Given that up to half of forestry practitioners in Europe are socially motivated \citep{diana_feliciano_decision_2025}, models that neglect social norms risk overlooking critical adoption dynamics and the spatial coordination of management decisions.

The stabilising influence of social norms reveals a crucial link to the system’s broader dynamics. By reinforcing prevailing behaviours, social interactions create feedbacks that not only foster spatial coherence but also give rise to path dependence and lock-in effects as explored in the next section.

\subsubsection{Critical Transitions and Path Dependence}
The critical transitions observed in the model illustrate a hallmark of complex adaptive systems: the coexistence of multiple stable regimes separated by sharp thresholds.
In the simulations, transitions between regimes occur once key behavioural thresholds are crossed, most notably when environmental attitudes strongly oppose the prevailing management practice. In addition, sufficient spatial diversity can enable the system to escape stabilisation near its initial configuration by allowing minority practices to spread and reshape long-term outcomes.

Beyond these thresholds, positive feedbacks between social norms, attitudes, and spatial clustering amplify small perturbations, reinforcing newly emerging configurations and making reversals difficult.

These reinforcing loops generate lock-in dynamics: configurations that align attitudes, social norms, and spatial patterns become self-stabilising, while alternative practices face high entry barriers.
As a result, returning the underlying drivers to their original values does not necessarily restore the previous equilibrium, a dynamic that mirrors hysteresis in other complex systems.

Similar lock-in phenomena have been observed empirically in agricultural and forestry systems, where established land management practices persist despite structural or policy changes \citep{chavez_path_2013, song_multifunctionality_2022, alpuche_alvarez_unraveling_2024}. Such path-dependent trajectories arise from self-reinforcing interactions between social, economic, and spatial structures — a dynamic long recognised in socio-spatial \citep{atkinson_urban_1996}, socio-ecological \citep{goldstein_unlocking_2023} and particularly agri-food \citep{conti_why_2021} systems.

In our model, this persistence arises through the interaction of social reinforcement and spatial path dependence: social influence fosters local convergence, which in turn produces spatial clusters that limit exposure to behavioural alternatives.
Once formed, these clusters maintain their internal coherence through normative pressure and reduced cross-boundary influence.

From a broader perspective, these findings suggest that behavioural interventions or policy incentives may have limited short-term effects if introduced into systems already stabilised by such feedbacks.
Effective transformation therefore requires identifying leverage points — such as increasing local heterogeneity, diversifying management practices, or weakening reinforcing social norms — that enable the system to cross critical thresholds toward more desirable equilibria.

\subsection{Reflection on Model Design}
\subsubsection{Conceptual and technical choices}
The behavioural model was guided by the Theory of Planned Behaviour \citep{ajzen_theory_1991}, selected as a suitable framework for integrating socio-psychological drivers into land managers' decision-making in ABLUMs. This choice was informed by the MoHuB framework \citep{schluter_framework_2017} and complemented by Bicchieri’s theory of descriptive social norms \citep{bicchieri_grammar_2005}, which provides an empirically grounded representation of peer influence through observed behaviour. The TPB was chosen because it incorporates socio-psychological determinants that recent empirical studies identify as central to land management decision-making \citep{swart_meta-analyses_2023, dessart_behavioural_2019}. Furthermore, it offers a high level of conceptual completeness and genericity \citep{schluter_framework_2017}, while remaining extendable \citep{ajzen_theory_1991}, making it operationalisable for agent-based modelling of environmental behaviour \citep{taraghi_integrating_2025}.

The implementation adopts a rule-based approach grounded in psychological theory to support explanatory clarity and theoretical transparency. Emergent dynamics can therefore be traced directly to explicit behavioural mechanisms. Modelling choices were guided by a balance between parsimony and realism, enabling transparent interpretation, replicability, and applicability across land use contexts. The framework can be parametrised for specific case studies where appropriate empirical data are available.

\subsubsection{Assumptions}
Building on the core premise that land managers’ decisions are shaped by socio-psychological factors instead of pure economic rationality, several key assumptions within the individual model components require consideration.

Environmental and productivist attitudes are represented as opposing but symmetric forces, positively correlated with extensification and intensification, respectively. Importantly, attitudes are modelled as personal biases, not as the strength of economic rationality; responsiveness to economic signals is separately controlled by the upper limit of the giving-in threshold. As shown in the results, this formulation does not imply a simple monotonic relation between attitudes and land management intensity, since their effects interact with social and economic drivers. Both, modelling assumptions and following emergent outcomes align with empirical observations of attitudes guiding land management decisions \citep{swart_meta-analyses_2023,breustedt_factors_2013, espinosa-goded_identifying_2013, grammatikopoulou_locally_2013}.

Furthermore, the formulation of behavioural inertia in our model assumes that large shifts in land management intensity require stronger economic stimuli, which aligns with empirical observations that change tends to occur incrementally \citep{lambin_land_2010, van_vliet_manifestations_2015}.

Social norms are modelled as descriptive — capturing what others do, shown to be relevant in empirical studies \citep{dessart_behavioural_2019,brown_empirical_2018,schmidtner_spatial_2012, lapple_spatial_2015}. While the model allows asymmetric influence for extensification and intensification, a symmetric setup was chosen here for simplicity.

\subsection{Limitations and Future Research}

The behavioural extension has been integrated and tested within a single agent-based land use model (CRAFTY). The proposed approach is conceptually generic and could be adapted to other land use models with modest refinements. For instance, models that rely on probabilistic adoption decisions (e.g. \citep{manson_modeling_2016, joffre_combining_2015,dou_land-use_2020}) could integrate the behavioural logic by transforming the giving-in threshold into an adoption probability function. Future work could explore how the extension performs across different agent-based land use models to help assess its transferability.

The model currently omits injunctive norms, which may lead to an underrepresentation
of moral or reputational expectations while emphasising empirical ones. Cultural values
such as emotional attachment to the land are also not explicitly included, though they
can be indirectly captured through behavioural inertia or transmitted via social norms. The modular structure of the framework, however, allows for straightforward extension to include additional normative and attitudinal dimensions in future work.

As a rule-based system, decision rules remain static during simulations, and agents do not develop new behavioural strategies. While this limits adaptive learning, it preserves interpretability in comparison to adaptive approaches (e.g. AI methods), which aligns with the model’s explanatory purpose. 

Environmental attitudes are treated as exogenous inputs rather than endogenously evolving states. This choice was made to maintain model simplicity and to give users flexibility in specifying behavioural scenarios. Nevertheless, the framework could be extended to include endogenous attitude dynamics, for example through opinion dynamics models or socio-ecological feedbacks that link land quality, ES provision, and land managers’ decision-making.
 
The present study consist of a stylised, theoretical assessment. A natural next step is to apply the integrated behavioural model in empirically grounded settings, such as large-scale applications of CRAFTY (e.g. CRAFTY-Europe \citep{brown_societal_2019}).

\section{Conclusion}
This study aimed to enhance land use modelling by developing a behavioural extension that explicitly integrates socio-psychological drivers, reflecting empirical insights into land managers’ decision-making. For demonstration purposes and to examine the emergent dynamics in isolation from confounding factors, the behavioural extension was integrated into an existing agent-based land use model and applied in stylised settings. The results show that introducing decision-making rules beyond economic rationality can substantially alter system-level land management intensity shares, landscape configuration, and ES provision. In particular, nonlinear feedbacks between behavioural factors, resource distribution, and demand levels give rise to complex and sometimes counter-intuitive emergent dynamics that can diverge from the direct decision rules and formal symmetries implemented at the agent level.

The inclusion of social networks further reveals how local interactions and social influence generate new spatial configurations and increase landscape connectivity. Positive behavioural feedbacks and diffusion through social networks promote path dependence, lock-in effects, and the emergence of multiple stable regimes. While the extension was tested within a single agent-based model, it is conceptually generic and transferable across modelling platforms with only minor adaptations. Future research can apply the integrated land use model with behavioural extension in empirically grounded settings.

\newpage
\renewcommand{\thefigure}{A\arabic{figure}}
\setcounter{figure}{0}
\renewcommand{\thetable}{A\arabic{table}}
\setcounter{table}{0}
\section*{Appendix A: Supplementary Figures}
\label{appendix_A}

This appendix provides supplementary figures supporting the model description, as well as additional analyses underpinning the results presented in the main text.
\begin{figure}[H]
    \centering
    \includegraphics[width=0.7\linewidth]{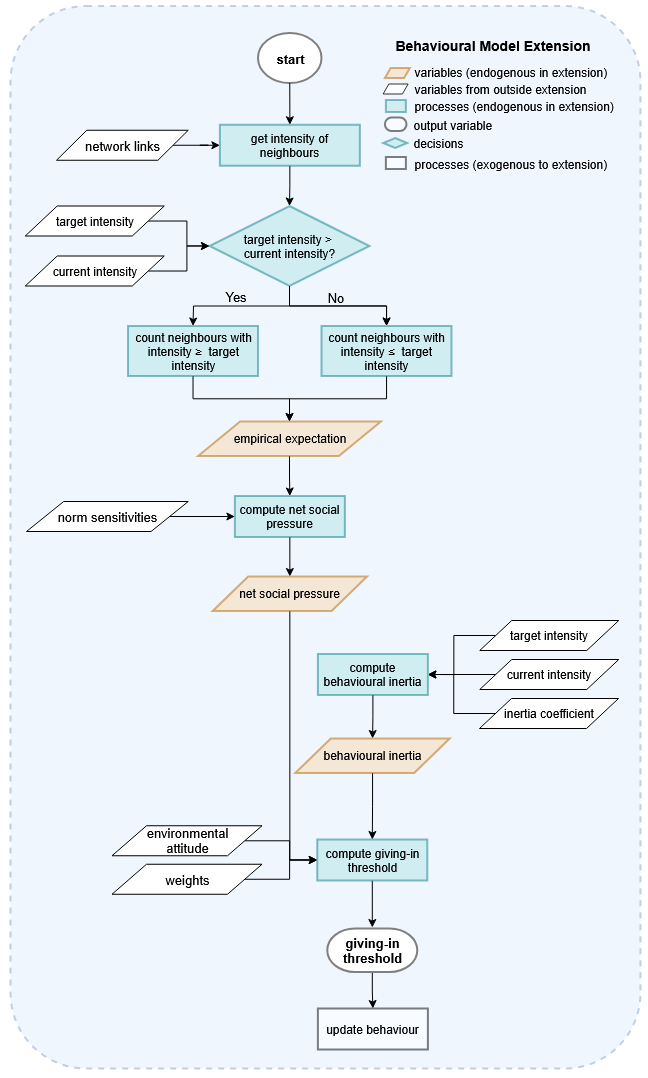}
    \caption{Flow diagram of the behavioural model extension.}
    \label{fig:flow_diagramm}
\end{figure}

\begin{figure}[H]
\centering
\includegraphics[width=0.5\linewidth]{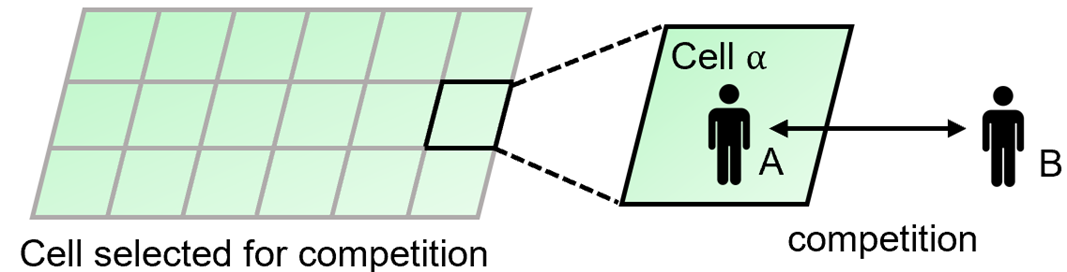}
\caption{Visualisation of the competition process within CRAFTY. }
\label{fig:competition}
\end{figure}

\begin{figure}[H]
    \centering
    \includegraphics[width=1\linewidth]{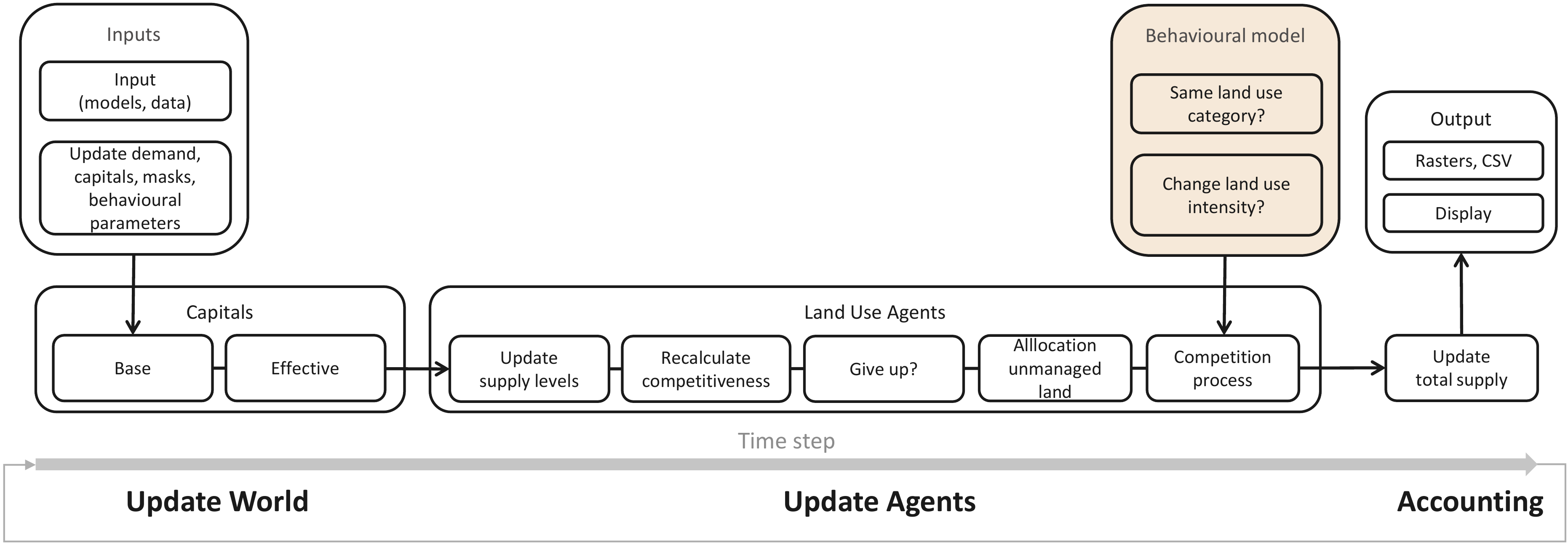}
    \caption{Sequence diagram of CRAFTY showing one time step with embedded behavioural model extension.}
    \label{fig:ext_CRAFTY_embedded}
\end{figure}

\begin{figure}[H]
\centering
\includegraphics[width=1\linewidth]{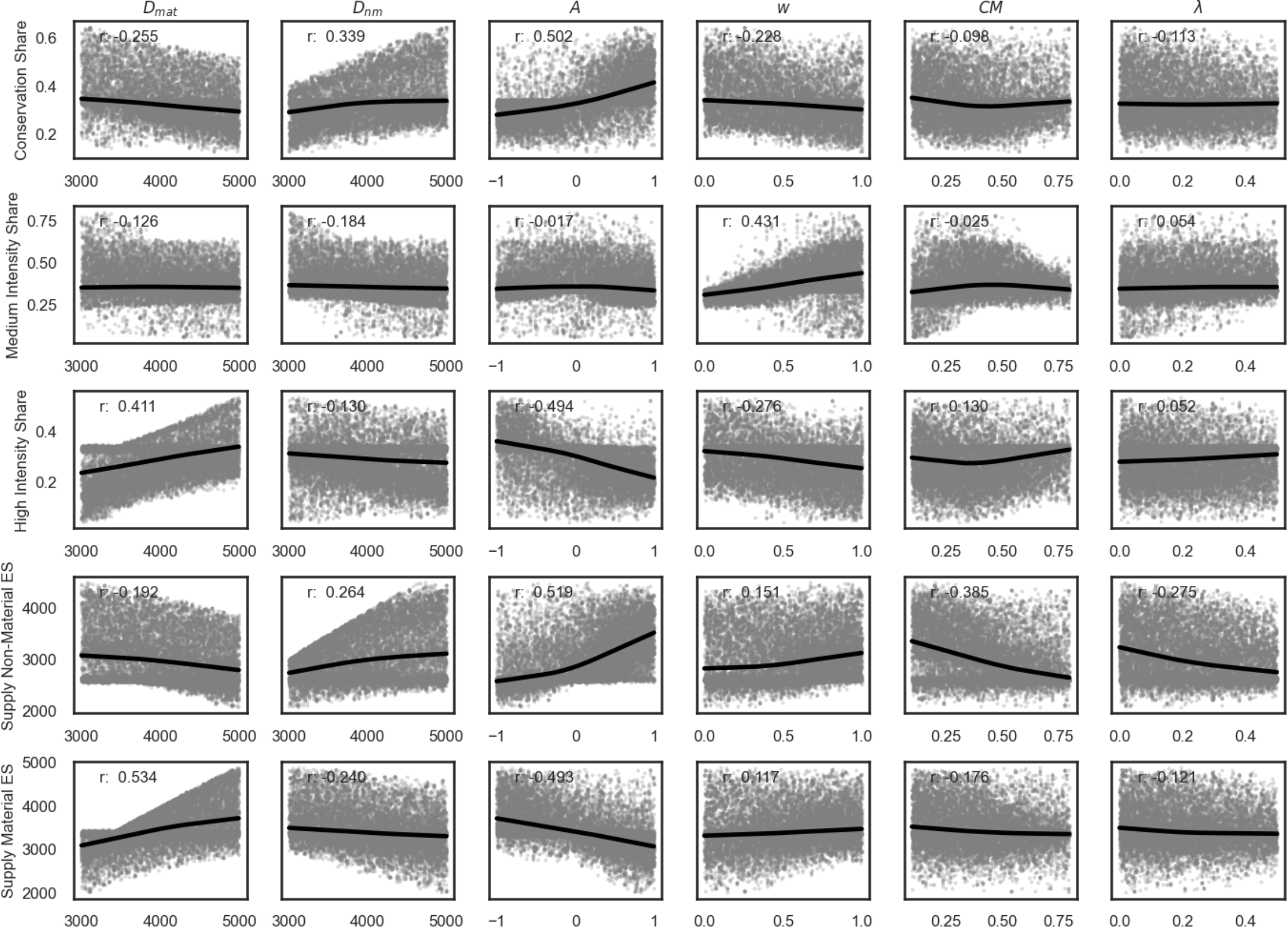}
\caption{Intensity shares and total supply of ES as a function of varied decision-making parameters and demand levels, based on the Sobol sensitivity analysis.}
\label{fig:SA_dependencies}
\end{figure}

\begin{figure}[H]
\centering

\subfigure[Conservation.]{
  \includegraphics[width=0.31\linewidth]{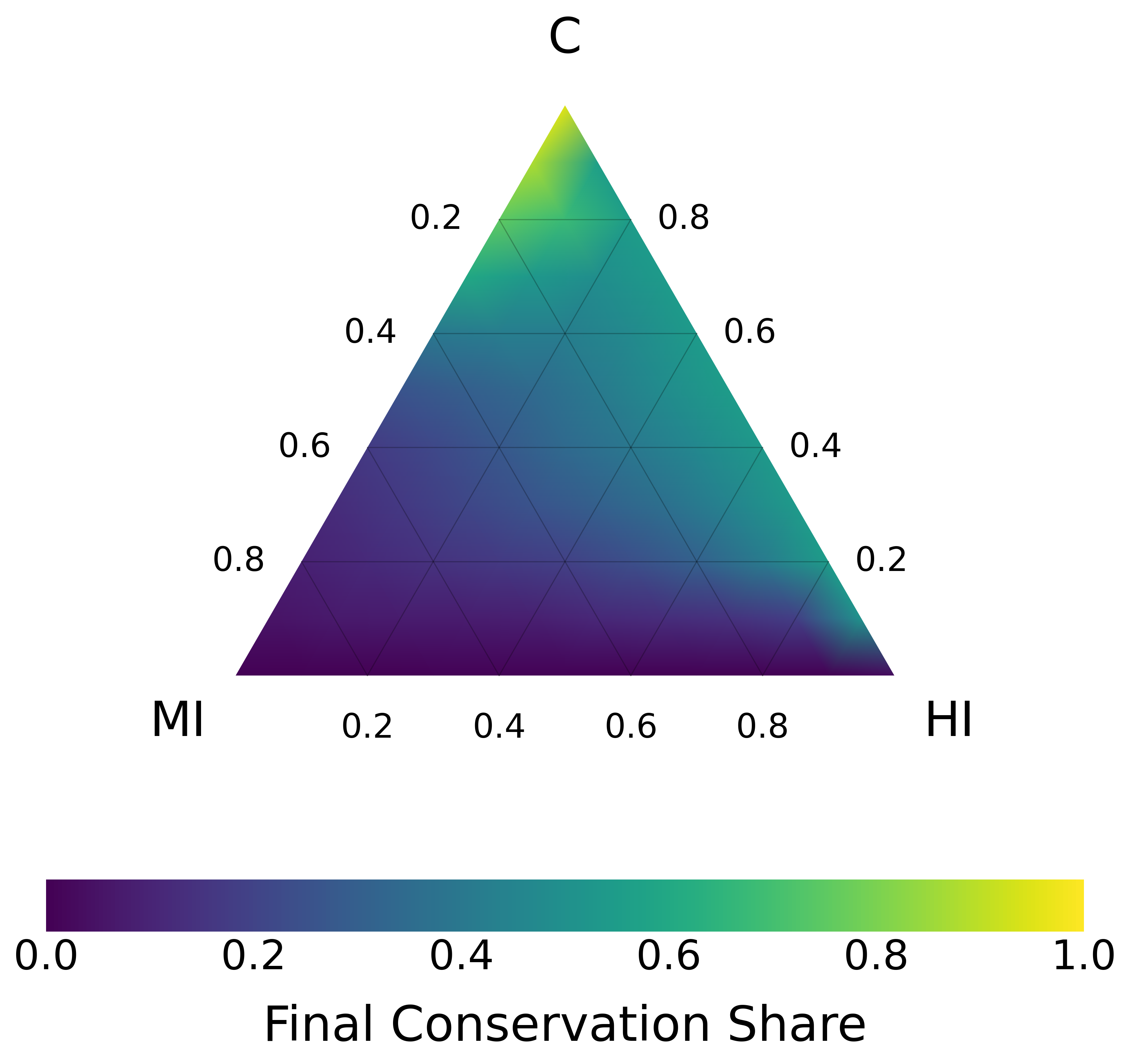}
  \label{fig:ternary_LIF}
}
\hfill
\subfigure[Medium intensity.]{
  \includegraphics[width=0.31\linewidth]{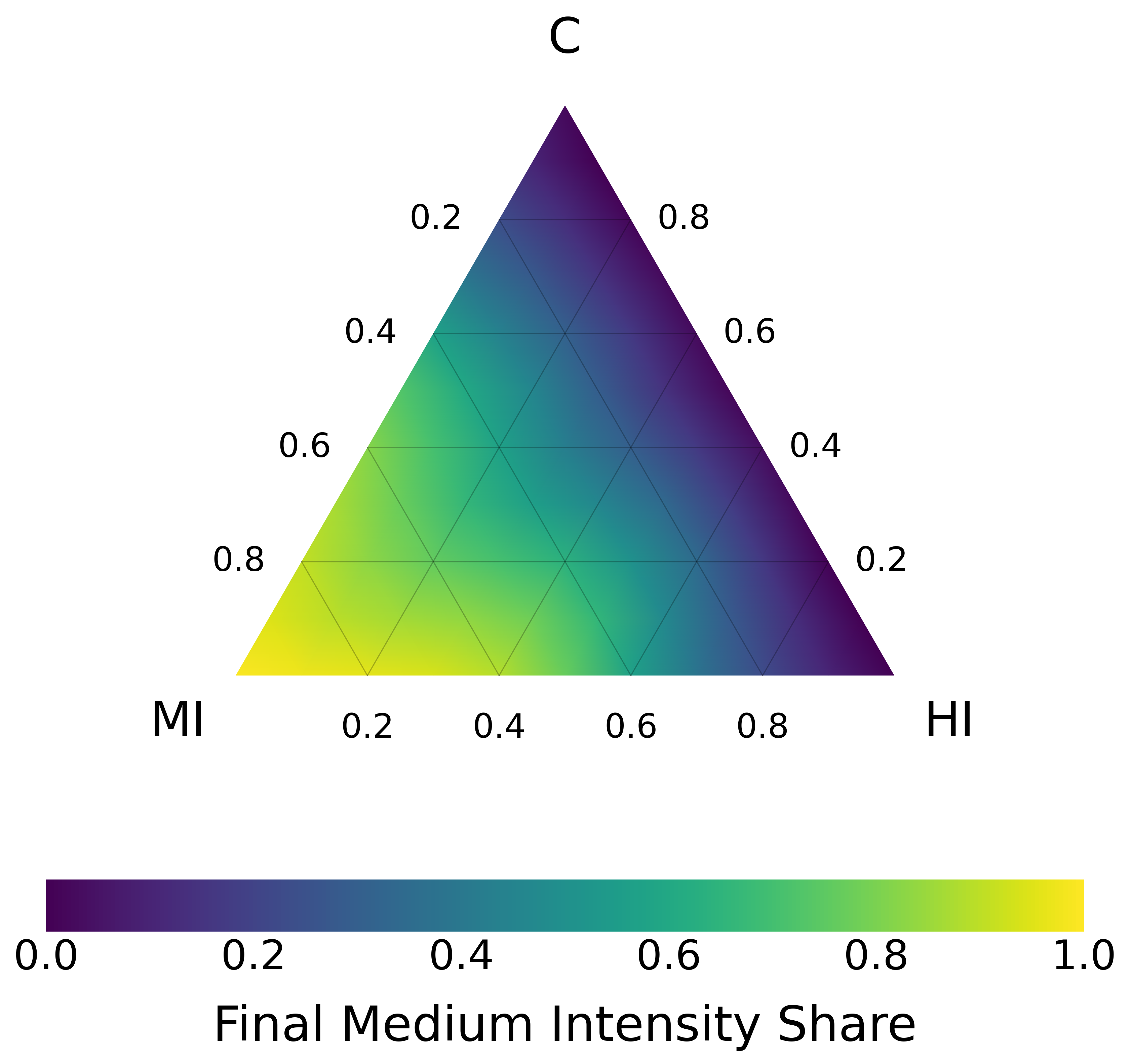}
  \label{fig:ternary_MIF}
}
\hfill
\subfigure[High intensity.]{
  \includegraphics[width=0.31\linewidth]{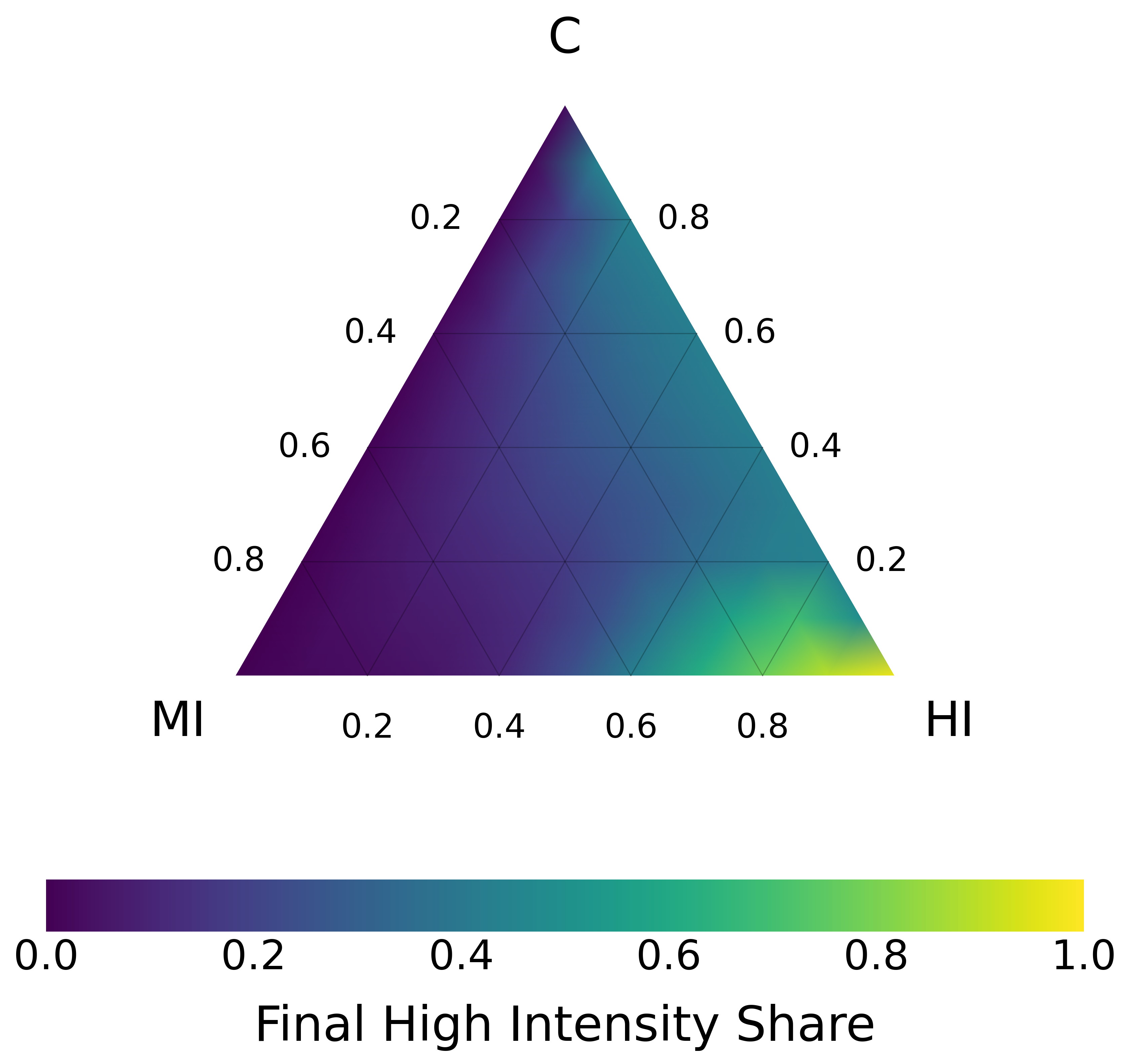}
  \label{fig:ternary_HIF}
}

\caption{Final land management intensity shares across all combinations of initial conditions.
Each corner of the ternary plots corresponds to a configuration where one land management practice dominates initially
(Medium Intensity (MI), High Intensity (HI), Conservation (C)).
Colour intensity indicates the relative share of that practice after model stabilisation.
Simulation parameters as listed in Table~\ref{tab:parameters_all}.}
\label{fig:ternary_final_shares}
\end{figure}

\begin{figure}[H]
    \centering
    \includegraphics[width=1\linewidth]{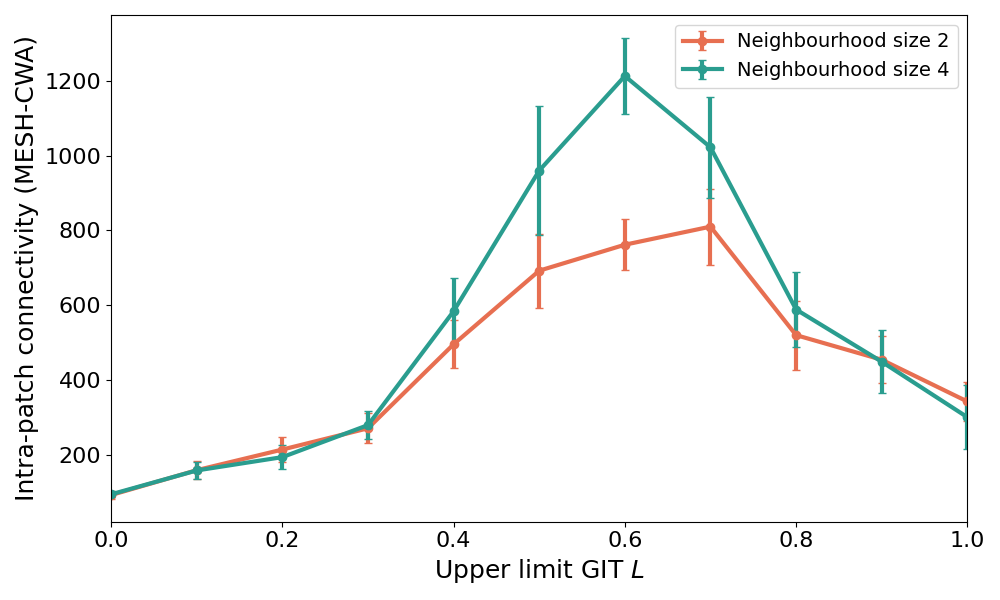}
    \caption{Intra-patch connectivity (MESH-CWA) in function of upper limit of the giving-in threshold ($L$). Simulation parameters as listed in Table~\ref{tab:parameters_all}.}
    \label{fig:connectivity_balanced}
\end{figure}

\begin{figure}[H]
    \centering
    \includegraphics[width=0.8\linewidth]{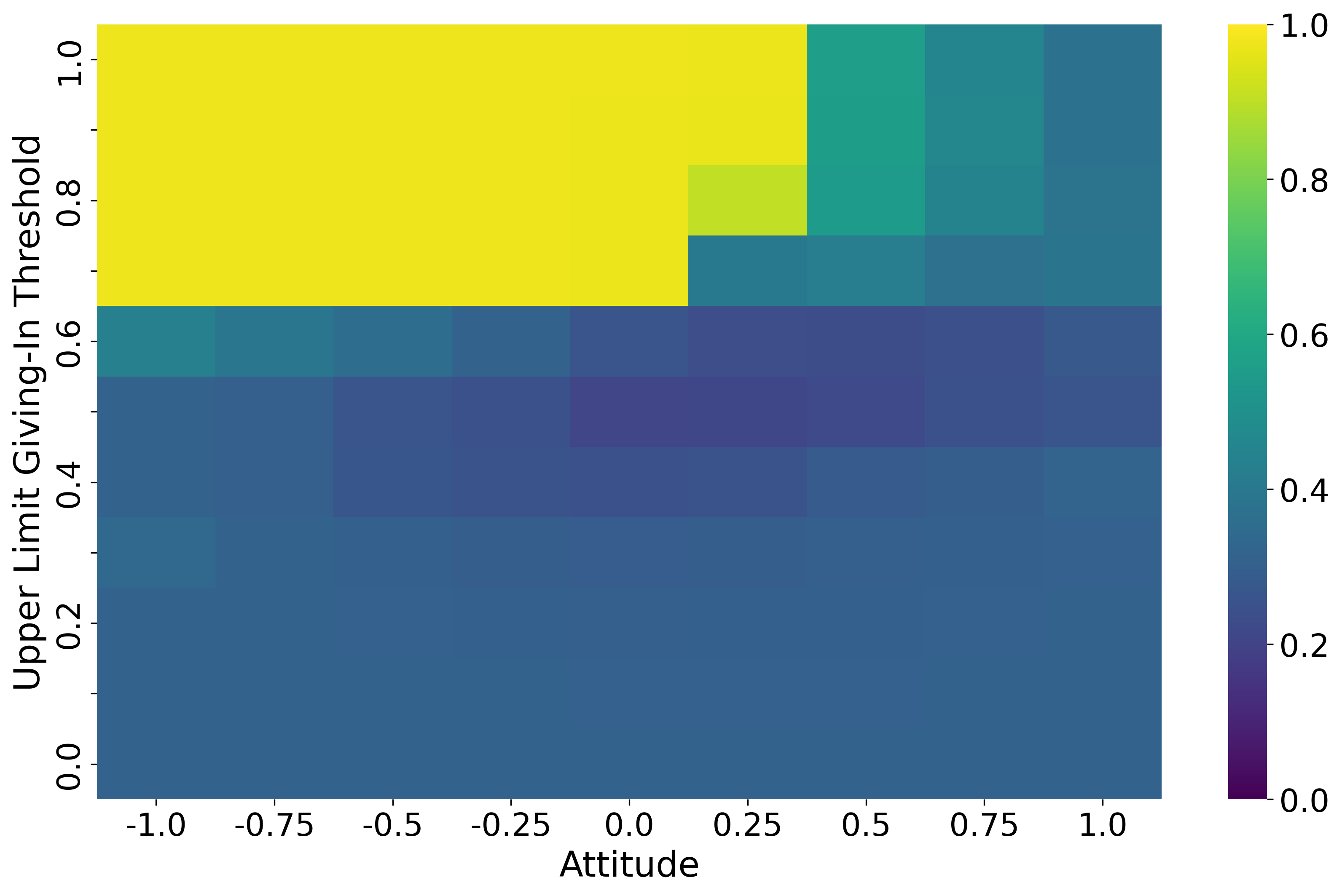}
    \caption{
    Final share of high intensity land management as a function of mean environmental attitude and the upper limit of the giving-in threshold.
    Colours indicate the relative dominance of high intensity land management after model stabilisation. Simulation parameters as listed in Table~\ref{tab:parameters_all}.
    }
    \label{fig:hif_heatmap_att_L}
\end{figure}

\begin{figure}[H]
\centering
\includegraphics[width=1\linewidth]{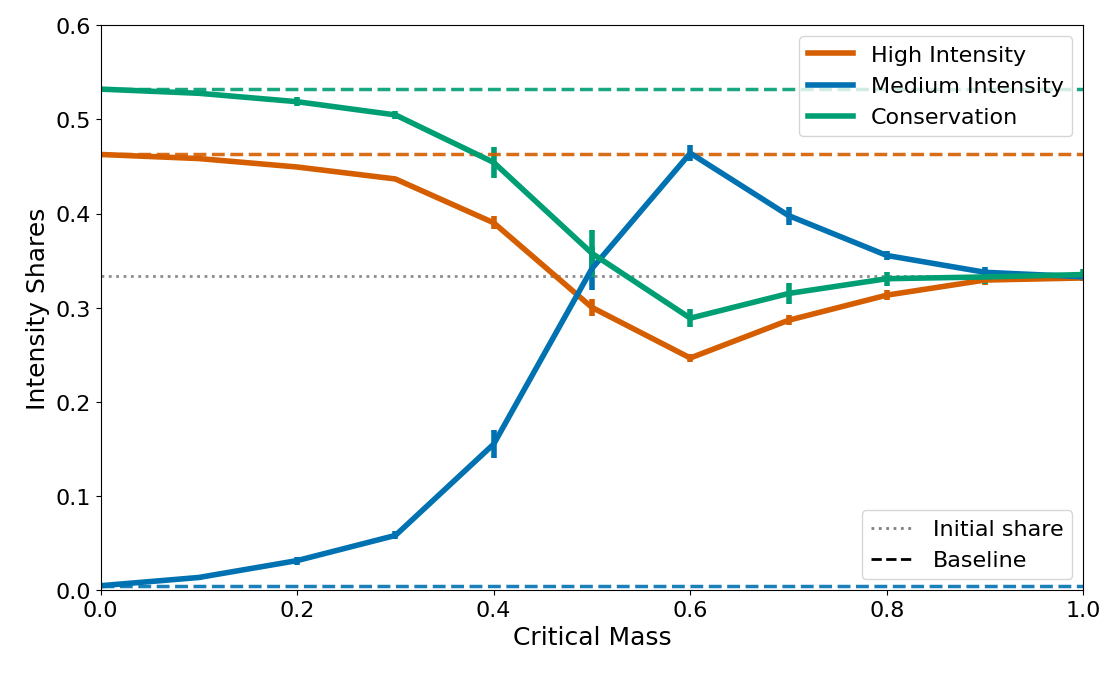}
\caption{Intensity shares after stabilization as a function of the critical mass threshold.
The dotted line marks the initial intensity shares (set equally across all land management practices). Dashed lines indicate the baseline dynamics of the purely economic model without social norms. Simulation parameters as listed in Table~\ref{tab:parameters_all}.}
\label{fig:critical_mass}
\end{figure}

\section*{Appendix B: Supplementary Tables}
This appendix provides tables defining the parameter configurations used in the sensitivity analysis and other experiments.
\begin{table}[H]
\centering
\caption{Parameter ranges used for the global Sobol sensitivity analysis.}
\label{tab:sobol_ranges}
\begin{tabular}{lcc}
\hline
\textbf{Parameter} & \textbf{Minimum} & \textbf{Maximum} \\
\hline
Environmental attitude $A$        & $-1$   & $1$ \\
Weight of social norms $w$         & $0$    & $1$ \\
Inertia coefficient $\lambda$      & $0$    & $0.5$ \\
Critical mass $CM_{\mathrm{int}}$, $CM_{\mathrm{ext}}$ & $0.1$ & $0.8$ \\
Demand material ES            & $3000$  & $5000$ \\
Demand non-material ES   & $3000$ & $5000$ \\
Neighbourhood radius $S_{nb}$      & $1$    & $5$ \\
Number of teleconnections $N_{tele}$ & $0$   & $2500$ \\
\hline
\end{tabular}
\end{table}

\begin{table}[H]
\centering
\caption{Parameter settings used in the different experiments (see corresponding figures for results). 
All parameters are spatially uniform and static across the landscape unless indicated as \textit{var.}, meaning they are systematically varied in the corresponding experiment.}
\label{tab:parameters_all}
\begin{tabular}{lccccccc}
\hline
\textbf{Parameter} & \textbf{Fig. \ref{fig:regimes}} & \textbf{Fig. \ref{fig:dyn_att}} & \textbf{Fig. \ref{fig:initial_conditions} \ref{fig:ternary_final_shares}} & \textbf{Fig. \ref{fig:critical_mass}} & \textbf{Fig. \ref{fig:connectivity_balanced}} & \textbf{Fig. \ref{fig:balanced_maps}}& Fig. \ref{fig:hif_heatmap_att_L} \\
\hline
$\lambda$ & 0 & 0 & 0 & 0 & 0 & 0& 0 \\
$w$ & 0.5 & var. & 1 & 1 & 1 & 1 & 0.5\\
$CM_{\mathrm{int}}$, $CM_{\mathrm{ext}}$ & 0.5 & 0.5 & 0.5 & var. & 0.5 & 0.5 & 0.5\\
$L$ & 0.65 & 1 & 1 & 1 & var. & 1 & var.\\
$D_{\mathrm{mat}}$, $D_{\mathrm{nm}}$ & 3500 & 3000 & 4000 & 5000 & 4000 & 4000 & 3500\\
$N_{\mathrm{tele}}$ & 0 & 0 & 0 & 0 & 0 & var.& 0\\
$S_{\mathrm{nb}}$ & 1 & 1 & 1 & 1 & var. & var. & 1\\
$k$ & 10 & 10 & 10 & 5 & 10 & 10 & 10\\
$A$ & var. & var. & -- & -- & -- & -- & var.\\
init. HI share & var. & 0.33 & var. & 0.33 & 0.33 & 0.33 & 1\\
init. MI share & var. & 0.33 & 0 & 0.33 & 0.33 & 0.33 & 0\\
init. C share & var. & 0.33 & var. & 0.33 & 0.33 & 0.33 & 0\\
\hline
\end{tabular}
\smallskip
\footnotesize Initial shares (init. HI, MI, and C share) represent the proportion of land managers 
initially operating high intensity, medium intensity, and conservation land use strategies, respectively 
(\(\sum = 1\)).
\end{table}

\newpage

\bibliographystyle{jasss}

\bibliography{references} 

\end{document}